\documentclass[onecolumn]{aastex7}

\usepackage{graphicx}
\usepackage{longtable}
\usepackage{booktabs}
\usepackage{graphicx}
\usepackage{subfig}

\begin{document}

\title{Probing Compact Objects in Wide-Orbit Binaries with Joint LAMOST LRS and MRS}
\shorttitle{LAMOST Compact Objects in Wide-Orbit Binaries}
\shortauthors{Liu et al.}

\author[0000-0002-2912-095X]{Hao-Bin Liu}
\affiliation{Department of Astronomy, Xiamen University, Xiamen, Fujian 361005, People's Republic of China}
\email{liuhaobin@stu.xmu.edu.cn}

\author[0000-0003-3137-1851]{Wei-Min Gu}
\affiliation{Department of Astronomy, Xiamen University, Xiamen, Fujian 361005, People's Republic of China}
\email[show]{guwm@xmu.edu.cn}

\begin{abstract}
Wide-orbit binaries serve as crucial laboratories for understanding stellar evolution and identifying quiescent compact objects. In this work, we search for compact objects in wide-orbit binaries by merging the LAMOST multi-epoch catalogs from LRS and MRS in the 12th data release. We specifically focus on sources with at least 20 observation epochs that clearly exhibit long-term radial velocity (RV) variations while remaining essentially stationary over short time scales. By constraining the mass function with Lomb-Scargle periods and RV ranges, we identified 74 single-lined spectroscopic binary candidates harboring potential compact objects with robust orbital solutions. These systems exhibit orbital periods ranging from 10 to 1000 days, with semi-amplitudes of velocity $K_1 \lesssim 50$~${\rm km\,s^{-1}}$ and mass functions $f(M_2)$ between 0.03 and 0.94~$M_{\odot}$. Combining $f(M_2)$ with SED-derived stellar parameters, we identify four strong compact object candidates with main-sequence companions (Class A), 9 systems likely consisting of either compact objects with giant/subgiant companions or mass-inverted Algol-type binaries (Class B), and 61 candidates with relatively lower mass ratios (Class C). Cross-matching with the \textit{Gaia} DR3 \texttt{nss\_two\_star\_orbit} catalog yields 16 sources, all of which exhibit orbital solutions consistent with our results. This study demonstrates the essential role of long-term spectroscopic monitoring in searching for compact objects in wide-orbit binaries and validating orbital solutions. The strategy of leveraging extended time baselines will be increasingly effective as spectroscopic databases continue to grow, enabling the systematic discovery of compact objects in wide orbits across the Galaxy.
\end{abstract}

\keywords{Compact objects (288)  --- Radial velocity (1332) --- Spectroscopic binary stars (1557)}

\section{Introduction} \label{sec:intro}
Binary systems containing compact objects, such as white dwarfs, neutron stars, or black holes, serve as fundamental laboratories for testing stellar evolution and binary interaction theories. While early investigations into these binaries were mainly shaped by interacting systems characterized by high-energy accretion \citep{1995cvs..book.....W, 2006ARA&A..44...49R, 2008LRR....11....8L, 2014MNRAS.438L..51M}, there is a growing consensus on the importance of studying the 'dormant' or quiescent population \citep{
2022NatAs...6.1203Y, 
2022ApJ...933..193Z, 2023MNRAS.518.1057E, 2023ApJ...954....4G,  2025PASP..137i4202N}. Recent advancements in high-precision astrometry, notably from the \textit{Gaia} mission, have successfully unveiled a number of wide-orbit compact object candidates \citep{2022arXiv220700680A,
2023A&A...677A..11G, 2023MNRAS.521.5927J, 
2024OJAp....7E..58E,
2025arXiv251005982M, 
Lam2026}. These non-accreting systems provide a more unbiased view of binary formation channels.

Complementary to astrometric efforts, multi-epoch spectroscopic surveys have emerged as a powerful tool for exploring this population across a broad range of orbital periods \citep{2018IAUS..330..350V, 2025arXiv251215904A}. In particular, the rich spectral library provided by the Large Sky Area Multi-Object Fiber Spectroscopic Telescope (LAMOST) facilitates a comprehensive search for these hidden objects using RV variations \citep{2020arXiv200507210L}.
Our previous work \cite[hereafter Liu24]{Liu2024} revealed 26 compact object candidates in single-lined binaries (SB1) using time-domain  spectra from the LAMOST medium-resolution survey (MRS). Subsequent studies have further characterized the nature of the compact object in two notable candidates \citep{2024ApJ...977..245Z, 2025JHEAp..45..428Z}. 
However, the Liu24 selection strategy relied primarily on significant RV variations of the visible companion, specifically focusing on those with an RV difference greater than 100~${\rm km\,s^{-1}}$. Although this approach efficiently identified compact object binaries, it was naturally biased toward short orbital period systems with large velocity amplitudes \citep{2017MNRAS.472.4497S}. As a result, our sample was largely concentrated in the short orbital period regime, with no targets having orbital periods longer than approximately 10 days. This leaves a vast, relatively unexplored parameter space for longer periods, where a significant population of compact objects may still be hidden.

In the present study, we combine LAMOST multi-epoch spectra from MRS and low-resolution survey (LRS), thereby substantially increasing the number of repeated observations for each target. With sufficient repeated measurements, we aim to systematically search for compact objects in long-orbital-period binary systems. The dense temporal coverage allows for well-sampled RV curves of the visible companions, improving sensitivity to systems with long orbital periods. This approach extends previous efforts by enabling exploration of long orbital period compact object candidates that were largely inaccessible in earlier searches. This paper is organized as follows. In Section~\ref{sec:selection}, we describe the data and the methods used to identify candidates. Section~\ref{sec:results} presents the main results and the classification of our sample. Finally, we summarize our conclusions and discussion in Section~\ref{sec:conclusion}.

\section{Method} \label{sec:selection}
\subsection{Data Reduction}
The MRS spectra have a spectral resolution of $R \sim 7500$, covering the blue wavelength range 4950--5350\,\AA\ and the red range 6300--6800~\AA, while the LRS spectra have a resolving power of $R \sim 1800$, spanning 3700--9000\,\AA. LAMOST's wide field-of-view and multi-fiber design enable large-scale spectroscopic surveys and provide repeated observations for a substantial number of targets \citep{2012RAA....12.1197C}. The LRS dataset contains approximately 12.6 million spectra obtained from 6,602 plates. The MRS dataset includes about 15.5 million spectra, comprising both single-epoch and coadd observations, and offers higher spectral resolution that is well suited for dynamical searches based on RV measurements.

We downloaded the multi-epoch spectroscopic catalogs from LAMOST DR12, including the LRS catalog \texttt{dr12\_v1.1\_LRS\_mec.fits.gz} and the MRS catalog \texttt{dr12\_v1.1\_MRS\_mec.fits.gz}. To construct a working sample suitable for dynamical searches of compact object binaries, we combined the two catalogs and focused on targets with sufficiently dense temporal sampling. A positional cross-match was performed between the two catalogs within a radius of 2\arcsec, allowing repeated observations obtained from both surveys to be associated with individual sources. For each matched source, we calculated the total number of spectroscopic epochs by combining both MRS and LRS measurements. In this study, the MRS data serve as the primary resource for identifying compact object candidates, while the LRS data provide complementary time-domain coverage.

The total number of observations is the first selection criterion. We required each target to have more than 20 spectroscopic epochs in total, combining MRS and LRS observations. This criterion ensures adequate temporal coverage for sampling RV variability, particularly for systems with long orbital periods. Applying this requirement to the combined LAMOST dataset reduced the initial parent sample to a smaller subset of approximately $5\times10^{4}$ sources. Additionally, no restrictions were imposed on any catalog parameters, such as spectral type or other stellar properties of the visible companion.

\subsection{Candidate Selection}

\begin{figure*}[t] 
    \centering
    \includegraphics[width=0.95\textwidth]{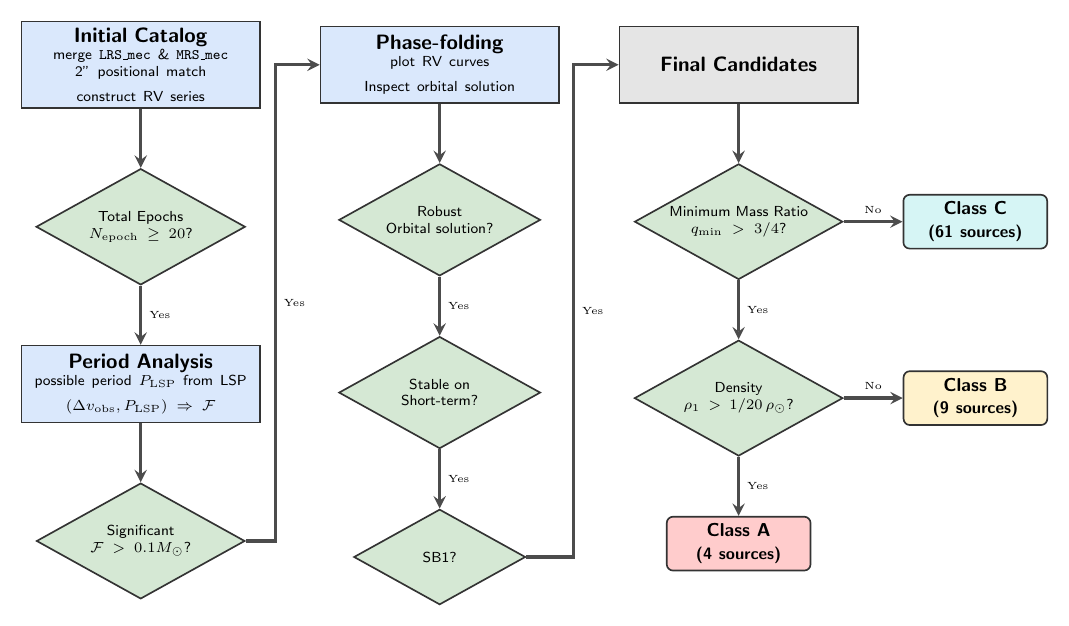} 
    \caption{Flowchart of the sample selection and classification process. }
    \label{fig:flowchart}
\end{figure*}

For each source in the combined catalog, we extracted the RV measurements provided by the LAMOST pipeline and constructed a RV series $\{v_{{\rm obs},i}\}$.
Owing to the large number of spectroscopic epochs available for each target, the orbital phase coverage is generally sufficient. As a result, the maximum observed RV difference $
\Delta v_{\rm obs} = \max(v_{{\rm obs},i}) - \min(v_{{\rm obs},i})$
can be used as an approximation to twice the RV semi-amplitude, i.e., $\Delta v_{\rm obs} \simeq 2K$, for the purpose of an initial dynamical assessment.

Using the Lomb--Scargle periodogram \cite[LSP,][]{1976Ap&SS..39..447L,1981ApJS...45....1S}, a preliminary period search was performed on the RV series. The analysis was carried out with the \texttt{LombScargle} class implemented in the \texttt{astropy.timeseries} module \citep{2013A&A...558A..33A,2018AJ....156..123A,2022ApJ...935..167A}, with the RV series and the corresponding observation times as inputs. Permitted by the multi-year temporal baseline of the data, the period search was optimized for long orbital periods. We set the \texttt{minimum\_frequency} to $1/10^{4}~\mathrm{day^{-1}}$ and the \texttt{maximum\_frequency} to $1~\mathrm{day^{-1}}$. This range was chosen to fully exploit the extended temporal coverage of the combined MRS and LRS observations. The highest peak in the LSP was adopted as the candidate period, denoted as $P_{\rm LSP}$. Based on the $P_{\rm LSP}$ obtained from the LSP, we define a mass-function-like screening parameter, denoted as $\mathcal{F}$, as:
\begin{equation}\label{eq:screening}
\mathcal{F} = 1.013\times10^{-7}
\left( \frac{\Delta v_{\rm obs}/2}{\mathrm{\rm km\,s^{-1}}} \right)^3
\left( \frac{P_{\rm LSP}}{\mathrm{day}} \right)~M_\odot.
\end{equation}
As a screening parameter, $\mathcal{F}$ is derived from the observed RV series, $\{v_{{\rm obs},i}\}$, and does not represent a true mass function. We applied a threshold on $\mathcal{F}$, selecting only sources with $\mathcal{F} > 0.1~M_\odot$, where this lower limit corresponds to a circular-orbit binary with an inclination of $90^\circ$ and a mass function of the secondary object equal to $0.1~M_\odot$. 

The RV series are folded using the corresponding $P_{\rm LSP}$ to construct phase-folded RV curves for remaining sources. 
Unreasonable RV curves are still produced by the severe deviation between the orbital period $P_{\rm orb}$ and  $P_{\rm LSP}$. 
Thus RV curves were inspected visually by examining the phase-folded RV curves for coherent orbital modulation. 
Such discrepancies are likely due to the concentration of observational epochs within limited orbital phases. 
Systems exhibiting clear and well-defined RV variations consistent with binary orbital motion were readily identified.

Non-uniform temporal sampling always carries the risk of misidentifying the orbital period \citep{2018ApJS..236...16V, 2010ApJ...722..937D}. 
In addition, We examined the original RV series for each candidate. 
The selected systems are required to show negligible RV variation within a single night or on short timescales ($\sim 10$ days), 
while exhibiting significant RV changes over longer timescales, consistent with the candidate long orbital periods. 
This procedure minimizes the likelihood that a system is misclassified as a long-period binary due to limited temporal sampling or short-term variations.

We further identify and exclude double-lined spectroscopic binaries (SB2s). For each source, we visually inspected the available MRS spectra, as they provide sufficient spectral resolution to identify SB2 \citep{2022MNRAS.517..356K,
2022ApJS..258...26Z, 
2023ApJS..266...18Z, 2024MNRAS.527..521K, 
2025ApJS..278...46G}. Specifically, the profiles of prominent stellar absorption lines were examined across multiple epochs. We focused on diagnostic absorption features, including H$\alpha$, \ion{Mg}{1}, \ion{Ca}{1}, and \text{Fe} lines, which are consistently covered within the LAMOST MRS wavelength ranges. 
We searched for characteristic SB2 signatures such as line splitting, asymmetric absorption profiles, or systematic changes in line shape between different observations, especially when the RV gets its maximum and minimum. Such features indicate the presence of two luminous stellar components contributing to the observed spectrum. 

\begin{figure*}[t] 
\centering \includegraphics[width=0.9\textwidth]{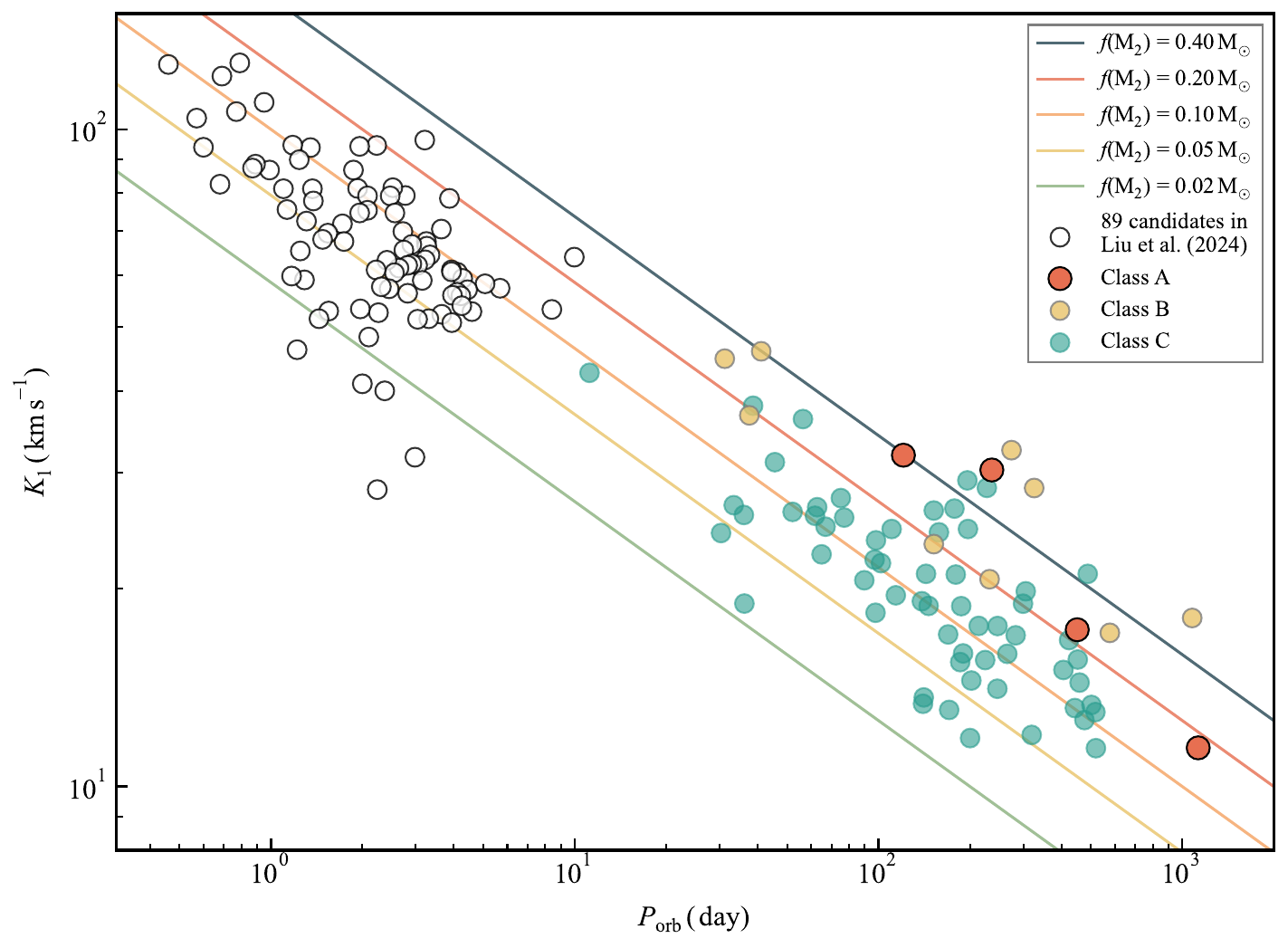}
\caption{
Orbital characteristics of the 74 candidates in the $K_1$--$P_{\rm orb}$ plane. Contours of constant mass function $f(M_2)$ (at $e=0$, from 0.02 to 0.40 $M_\odot$) are shown as colored lines. Filled circles indicate our candidates, classified into Class~A (red), B (yellow), and C (teal) according to the criteria described in Section \ref{sec:classification}. These markers are consistent with Figure~\ref{fig:fm_density}. We show candidates in Liu24 (open circles) alongside our new candidates. Notably, the Liu24 candidates are concentrated at significantly shorter orbital periods, whereas our sample extends to longer orbital periods. }
\label{fig:sample} \end{figure*}

\section{Results} \label{sec:results}

We selected 74 promising compact object candidates in binaries. These candidates exhibit coherent RV variations consistent with single-lined spectroscopic binary motion and show no detectable signatures of a luminous secondary object in their spectra.
As shown in Figure~\ref{fig:sample}, these are promising candidates for the combination of long orbital periods and moderate velocity amplitudes. This population occupies a distinct parameter space compared to the shorter-period systems reported in Liu24, thereby extending the search for compact objects into the long-period regime.

The orbital parameters of the 74 selected candidates cover a broad range of the parameter space, consistent with a population of wide-separation binaries. The candidate orbital periods are distributed between approximately $10^{1}$ and $10^{3}$ days. We did not prioritize systems with shorter orbital periods. Conversely, the identification of systems with even longer orbital periods is currently constrained by the total duration of the observational baseline. The observed RV semi-amplitudes for these candidates typically range from 10--50~${\rm km\,s^{-1}}$. 

\subsection{Template Matching}

For each of the 74 selected candidates, we measured RVs from LAMOST spectra using theoretical stellar spectra templates. For every target, we prioritized the medium-resolution spectrum with the highest signal-to-noise ratio to serve as the primary reference. We combined the coadded data from the blue and red arms into a single spectral vector to maximize the available spectral features for template matching.

The matching process was performed utilizing the MARCS synthetic spectral grid in the software \texttt{StellarSpecModel} \citep{stellarSpecModel2025}. Three stellar atmospheric parameters (effective temperature, surface gravity and metallicity) were free parameters for the grid interpolator. In addition, rotational velocity and RV are incorporated in the template to ensure the best possible match with the observed line profiles. We employed the \texttt{Spectool} software package \citep{spectool2025} to measure the rotational velocities, following a methodology similar to that described in \citet{2025ApJ...986...34Z}.  To determine the optimal parameter set, we defined the logarithmic likelihood function as follows:

\begin{equation}
\ln \mathcal{L} = -\frac{1}{2} \sum_{i=1}^{N} \left[ \frac{(F_{i} - F_{{\rm m},i}(\theta))^2}{\sigma_{i}^2 + \sigma_{\text{sys}}^2} + \ln(2\pi(\sigma_{i}^2 + \sigma_{\text{sys}}^2)) \right],
\end{equation}
where $F_i$ represents the observed normalized spectral flux, $F_{{\rm m},i}(\theta)$ is the synthetic flux generated by the MARCS interpolator for the parameter vector $\theta$, and $\sigma_i$ is the statistical uncertainty. The systematic error $\sigma_{\text{sys}}$ is incorporated into the total uncertainty to account for potential discrepancies between the synthetic models and observed spectra.
The posterior probability density was sampled using the Markov Chain Monte Carlo (MCMC) ensemble sampler \texttt{emcee} \citep{foreman2013emcee}.

\begin{figure*}[t] 
\centering \includegraphics[width=0.9\textwidth]{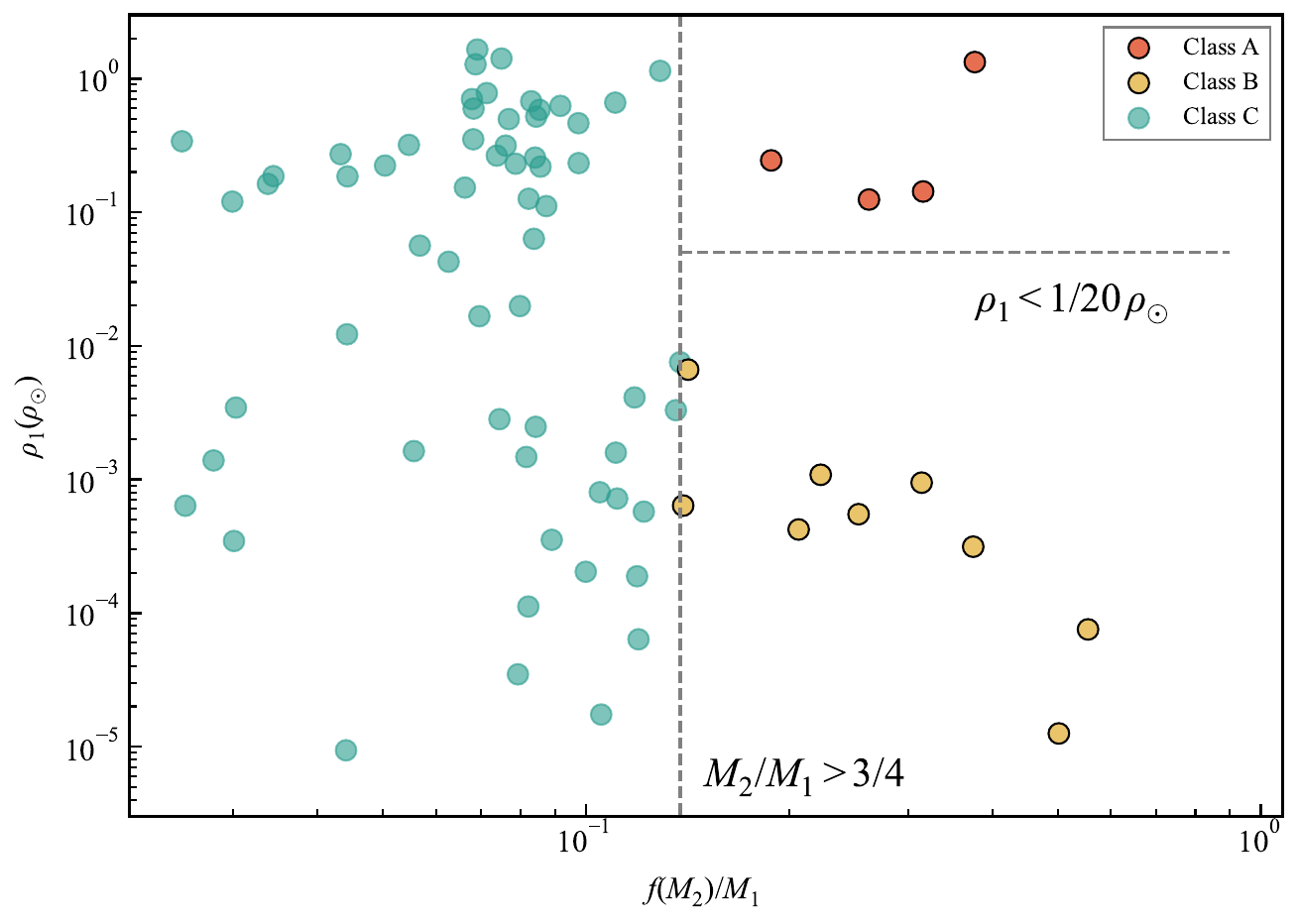} 
\caption{
Distribution of the mass function ratio $f(M_2)/M_1$ versus the mean density of visible companion $\rho_1$ for our sample of 74 candidates. 
We classify our candidates into three distinct categories based on their mass functions, mass ratios, and the evolutionary status of the primary stars. Class~A (orange) represents systems with mass functions exceeding $0.138\,M_1$ (corresponding to a mass ratio $q > 3/4$) and a primary star mean density $\rho_1 \approx M_1 R_1^{-3} < 1/20\,\rho_\odot$; these criteria specifically isolate systems consisting of a potentially compact object and a main-sequence companion. Class~B (yellow) includes targets that exhibit high mass ratios similar to Class~A but are characterized by visible primaries in the post-main-sequence stage, such as subgiants or giants, as inferred from their atmospheric parameters. Finally, Class~C (teal) comprises the remaining targets with lower mass functions ($f(M_2)/M_1 \le 0.138$), typically associated with secondaries significantly less massive than their primaries.}
\label{fig:fm_density} \end{figure*}

\subsection{Orbital Solutions}

We measured the RVs for each epoch across our 74 candidates using the resolution-matched templates. For the MRS, the coadded data from both the blue and red arms were integrated into a single spectral string before performing the cross-correlation. For the LRS spectra, the observed flux was directly cross-correlated against the synthetic templates. To ensure accuracy, all MARCS templates were explicitly downgraded to match the specific instrumental resolution of each observation (either MRS or LRS) prior to the cross-correlation function (CCF) measurement. Then, RVs were derived from the CCF module of Spectool. All uncertainties presented in this work were estimated through Monte Carlo simulations and correspond to the $1\sigma$ confidence level. RVs, along with the derived uncertainties, are summarized in Table~\ref{tab:rv}.

To characterize the dynamical properties of these systems, we fitted the resulting time-series RVs with a Keplerian model. The model predicts the RV at a given time $t$ by numerically solving the Kepler equation $M = E - e \sin E$, where $M = 2\pi(t - t_0) / P_{\rm orb}$ is the mean anomaly and $E$ is the eccentric anomaly. We implemented a numerical solver (via \texttt{scipy.optimize.fsolve}) to determine $E$, which was subsequently used to calculate the true anomaly $\theta$ through the relation $\tan(\theta/2) = \sqrt{(1+e)/(1-e)} \tan(E/2)$. The complete RV model is expressed as:
\begin{equation}
    V_{\rm r}(t) = K_1 \{ \cos[\theta(t) + \omega] + e \cos\omega \} +  \gamma\,,
\end{equation}
where $\theta(t)$ is calculated as described above, and the model includes six free parameters: the systemic velocity $\gamma$, representing the motion of the binary's barycenter relative to the observer; the RV semi-amplitude of the primary star $K_1$; the orbital period $P_{\rm orb}$; the eccentricity $e$, which defines the shape of the orbit; the argument of periastron $\omega$, specifying the orientation of the orbit in the orbital plane; and the time of periastron passage $t_0$. We implemented a MCMC framework to sample the posterior distributions of the orbital elements. The best-fit values are adopted from the medians of the resulting distributions, with the associated $1\sigma$ uncertainties derived from the $16^{th}$ and $84^{th}$ percentiles. These values are listed in Table~\ref{tab:orbit}.

The mass function is related to the masses of the two components and the orbital inclination, which is defined as follows:
\begin{equation} \label{eq:mass_function}
f(M_2) = \frac{M_2^3 \sin^3 i}{(M_1 + M_2)^2}
       = \frac{K_1^3 P_{\rm orb}}{2\pi G}\,(1-e^2)^{3/2},
\end{equation}
where $M_1$ is the mass of the optically visible star; $M_2$ is the mass of the unseen object; $i$ is the inclination, and $G$ is the gravitational constant.
Utilizing the best-fit orbital parameters, we calculated the mass function $f(M_2)$ for each system, which serves as a fundamental dynamical constraint on the nature of the unseen object.

\subsection{Classification of Candidates}

\label{sec:classification}
To obtain robust stellar parameters, we performed spectral energy distribution (SED) modeling of the observed multi-band photometry using the publicly available SED fitting code \texttt{ARIADNE} \citep{Vines_Jenkins2022}.
\texttt{ARIADNE} utilizes a Bayesian model-averaging approach to match theoretical atmosphere grids.
By integrating the \textit{Gaia} DR3 parallax as a prior to constrain the distance, we derived key stellar parameters including effective temperature ($T_{\rm eff}$), surface gravity ($\log g$), metallicity ([Fe/H]), and stellar radius ($R_1$). These parameters, along with the calculated bolometric luminosity, were subsequently interpolated onto the MESA Isochrones \& Stellar Tracks \cite[MIST;][]{dotter2016,choi2016,paxton2011,paxton2019} stellar models to estimate the mass of the visible primary star ($M_1$).

Based on the derived stellar and orbital parameters, we classified the 74 candidates into three distinct groups as illustrated in Figure~\ref{fig:fm_density}. 
The classification is primarily based on  the stellar density $\rho_1$ and the mass-function ratio, defined as $f(M_2)/M_1$. 
From the definition of the mass function (Equation \ref{eq:mass_function}), we can derive the following relationship:
\begin{equation} \label{eq:mass_function_ratio}
\frac{f(M_2)}{M_1} = \frac{q^3 }{(1 + q)^2}\,\sin^3 i,
\end{equation}
where $q \equiv M_2/M_1$ is the mass ratio. 
At a given orbital inclination, $f(M_2)/M_1$ increases with $q$.
We adopted a threshold of $f(M_2)/M_1 = 0.138$ to identify systems with large mass ratios. Specifically, this threshold corresponds to a minimum mass ratio of $q_\mathrm{min} = 3/4$, which is reached at $i = 90^\circ$. Systems with $f(M_2)/M_1 < 0.138$ are categorized as Class C (teal circles in Figure~\ref{fig:sample} and \ref{fig:fm_density}), representing low mass ratio candidates.

For systems exceeding this mass ratio threshold, we find that the stellar mean density, defined as $\rho_1 = (M_1/M_{\odot})(R_1/R_{\odot})^{-3}\,\rho_{\odot}$, provides a naturally clear separation between the populations in our specific sample. Specifically, a threshold of $\rho_1  = 1/20\,\rho_{\odot}$ effectively distinguishes candidates with main-sequence companions from those with more evolved primaries. Candidates with $\rho_1 >1/20\,\rho_{\odot}$ are designated as Class A (orange circles in Figure~\ref{fig:sample} and \ref{fig:fm_density}), while those with lower densities are labeled Class B (yellow circles in Figure~\ref{fig:sample} and \ref{fig:fm_density}). The mean density serves as a indicator of the primary's evolutionary stage, effectively distinguishing between main-sequence stars and more evolved subgiant or giant stars. As shown in Table~\ref{tab:params}, the $\log g$ values for several giants are reported to exceed $4.0$ dex. These values would imply unphysically high gravitational masses for such evolved objects, thereby highlighting the significant errors inherent in the $\log g$ measurements for our candidates. Consequently, we adopt the mean density to naturally delineate the main-sequence and evolved populations.

Given that $q > 3/4$ for these systems, if the secondary were a normal main-sequence star, its luminosity would be sufficient to produce detectable spectral lines or significant continuum contributions in our LAMOST spectra. The absence of such secondary signals in our high-S/N observations strongly suggests that the unseen objects in Class A systems are likely compact objects.
One of our Class A candidates, J060228.6+280758.3, has been firstly identified as a binary consisting of a G-type main-sequence star and a compact object \citep{2024ApJ...964..101Z}. While specific mass estimates may vary depending on the choice of stellar evolution models, their findings consistently point to a high-mass compact object (exceeding $1.36~M_\odot$).

For Class B systems, although the ratio $f(M_2)/M_1$ is relatively large, the relatively lower stellar densities suggest that the visible companions are likely evolved stars, such as subgiants or giants. In these cases, the high luminosity of an evolved companion could potentially mask the spectral signatures of a main-sequence star. Class B targets could be identified either as compact object binary candidates or as post-mass-transfer Algol-type binaries. In the latter scenario, these systems have already undergone mass ratio reversal, in which the observed low-mass giant is the stripped remnant of the former primary. Despite their evolutionary history, they currently exhibit no significant active mass transfer.

Class C systems, characterized by $f(M_2)/M_1 < 0.138$, represent the most ambiguous category in our sample. Due to the low mass function relative to the primary mass, the gravitational influence of the secondary is insufficient to distinguish between a low-mass main-sequence star, a white dwarf, or a high-mass object viewed at a very low orbital inclination. Consequently, while these systems are included in our survey, they remain lower-priority candidates for compact object searches without further constraints from multi-epoch high-resolution spectroscopy or astrometric data.

It is possible that these massive yet dim secondaries could be close pairs of low-mass stars. Such triples can sometimes be detected through low-amplitude eclipses in light curves. We checked the available TESS light curves, and none of our candidates show clear evidence of being a triple system. While several sources exhibit quasi-periodic variations with periods of approximately 5--10 days, we cannot rule out contamination from nearby starlight because the signals are neither distinct nor stable. Additionally, one target in Class~C, J230706.97+352452.3, exhibits an EA-type light curve corresponding to its $P_{\rm orb}$, indicating it is a binary star system rather than a compact object candidate. Further verification could rely on high-resolution spectroscopy to detect potential tertiary spectral signatures.

\subsection{\textit{Gaia} Astrometry}

\begin{table*}[t]
\centering
\caption{\textit{Gaia} DR3 Cross-match Results for a Subset of Candidates}
\label{tab:gaia_info}
\footnotesize
\begin{tabular*}{\textwidth}{@{\extracolsep{\fill}}lcrcrrc}
\hline
\hline
LAMOST designation & \textit{Gaia} DR3 source id & \texttt{plx} & \texttt{ruwe} & $P_{\rm orb}$ & $P_{\rm gaia} $ & NSS solution type \\
 & & (mas) & & (days) & (days) & \\

\hline
J084423.83+152520.8 &          609651611028044544 &         1.67 &          4.14 & $450.9 \pm 16.9$ & $451.9 \pm 2.6$ &              Orbital \\
J055223.94+282946.1 &         3443195600969981440 &         0.38 &          1.27 &   $56.4 \pm 0.1$ &  $56.4 \pm 0.1$ &                  SB1 \\
J091705.56+415450.3 &          816137661341669760 &         1.98 &          1.16 &  $110.5 \pm 0.1$ & $110.2 \pm 0.4$ &                  SB1 \\
J142728.19+453124.5 &         1494473388638937728 &         1.04 &          1.79 &  $519.5 \pm 0.9$ & $527.4 \pm 5.3$ &                  SB1 \\
J142728.19+453124.5 &         1494473388638937728 &         1.04 &          1.79 &  $519.5 \pm 0.9$ & $511.2 \pm 5.0$ &              Orbital \\
J230706.97+352452.3 &         1914890969694599296 &         1.01 &          1.09 &   $11.2 \pm 0.1$ &  $11.2 \pm 0.1$ &                  SB1 \\
J044741.50+483453.0 &          255981326450428160 &         0.77 &          1.49 &  $282.8 \pm 0.2$ & $284.5 \pm 1.1$ &                  SB1 \\
J060228.55+280758.8 &         3431326755205579264 &         2.92 &          1.96 &  $120.8 \pm 0.2$ & $120.9 \pm 0.1$ &                  SB1 \\
J041014.93+533012.6 &          275234290455134336 &         0.51 &          1.04 &  $158.1 \pm 0.1$ & $158.0 \pm 0.4$ &                  SB1 \\
J034116.64+492438.1 &          249599279929183744 &         0.43 &          1.22 &  $488.2 \pm 0.4$ & $487.1 \pm 1.1$ &                  SB1 \\
J035243.07+251621.9 &           66896456396598400 &         3.36 &          5.35 &  $224.4 \pm 0.1$ & $222.5 \pm 0.9$ &      AstroSpectroSB1 \\
J084437.03+194239.1 &          661148268907314432 &         5.76 &          6.38 &  $143.4 \pm 0.1$ & $142.8 \pm 0.2$ &      AstroSpectroSB1 \\
J030035.16+560301.8 &          459792402416078080 &         0.45 &          2.34 &  $517.5 \pm 2.7$ & $518.1 \pm 9.8$ &                  SB1 \\
J084925.85+111657.0 &          598890862525288192 &         3.79 &          2.37 &  $114.0 \pm 0.1$ & $114.7 \pm 0.2$ &                  SB1 \\
J031324.69+545919.5 &          447622629760600960 &         0.49 &          1.61 &  $274.0 \pm 0.2$ & $294.2 \pm 0.9$ &                  SB1 \\
J050636.51+483830.8 &          255290523910817024 &         0.15 &          1.02 &  $227.3 \pm 0.1$ & $230.1 \pm 1.7$ &                  SB1 \\
J114552.77+351726.6 &         4031997035561149824 &         1.36 &          1.27 &  $138.9 \pm 0.1$ & $140.1 \pm 0.4$ &                  SB1 \\
J084314.39+133049.6 &          609120272034206208 &         1.89 &          6.14 & $1127.9 \pm 3.5$ & ... & SecondDegreeTrendSB1 \\
\hline
\end{tabular*}
\begin{flushleft}
\textbf{Notes.} $P_{\rm orb}$ is the orbital period derived from our LAMOST RV analysis. $P_{\rm gaia}$ is the period from the \textit{Gaia} DR3 NSS catalog. Plx and RUWE are collected from the main \textit{Gaia} source catalog. \texttt{SB1}, \texttt{Orbital}, and \texttt{AstroSpectroSB1} solutions are retrieved from the \texttt{nss\_two\_body\_orbit} table, while the \texttt{Trend} (SecondDegreeTrendSB1) solution is obtained from the \texttt{nss\_non\_linear\_spectro} table.
\end{flushleft}
\end{table*}

\textit{Gaia} DR3 \citep{2023A&A...674A...1G} provides the most extensive census of binary stars 
compiled by the astronomical community to date, delivering orbital solutions 
derived from astrometric and/or spectroscopic measurements for 
approximately $8.1 \times 10^5$ stars.
\citep{Halbwachs2022, Holl2022}. 
We cross-matched our 74 candidates with the \textit{Gaia} DR3 Non-Single Star (NSS) catalog. There are 16 candidates with available \textit{Gaia} orbital solutions, as summarized in Table~\ref{tab:gaia_info}. For these overlapping sources, the orbital periods derived from our LAMOST time-series RVs are in good agreement with those reported by \textit{Gaia}. According to the \textit{Gaia} \texttt{nss\_solution\_type}, the majority of these systems are classified as \texttt{SB1} (spectroscopic binaries with a single-lined solution), while four candidates possess astrometric orbital solutions. 

For the four targets with available astrometric solutions, we derived their orbital inclinations $i$ from Thiele-Innes elements. Combining inclinations with  mass functions $f(M_2)$ and isochrone-based primary masses $M_1$, we estimates the masses of the secondary $M_2$. Specifically, J084423.83+152520.8 has a astrometric inclination of $64.6^\circ \pm 6.3^\circ$, yielding a secondary mass of $0.97 \pm 0.12\,M_\odot$. For J142728.19+453124.5, J035243.07+251621.9, and J084437.03+194239.1, the derived inclinations are $80.1^\circ \pm 8.0^\circ$, $61.8^\circ \pm 4.8^\circ$, and $83.9^\circ \pm 1.6^\circ$, respectively, which correspond to secondary masses of $0.59 \pm 0.02\,M_\odot$, $0.59 \pm 0.01\,M_\odot$, and $0.78 \pm 0.01\,M_\odot$. 

One of Class A candidates, J084314.4+133049.9, is classified with a \texttt{SecondDegreeTrendSB1} solution in the \texttt{gaiadr3.nss\_non\_linear\_spectro} catalog, further characterized by a significantly high RUWE of approximately 6.14. Such an elevated RUWE value underscores the fact that the astrometric measurements are poorly described by a single-star model, strongly suggesting the presence of an unresolved object. The identification of an acceleration solution for this source provides critical independent support for our derived long-period orbital solution ($P_{\rm orb} \approx 1127.9$~days); since the orbital period substantially exceeds the \textit{Gaia} DR3 observational baseline, the astrometric signature naturally manifests as a non-linear acceleration trend rather than a closed orbital solution.

\section{Conclusions and Discussion}\label{sec:conclusion}
In this work, we searched for compact objects in wide-orbit binaries by merging the LAMOST DR12 LRS and MRS multi-epoch catalogs. We identified 74 binary candidates with robust orbital solutions from a sample of sources with at least 20 observation epochs. These systems exhibit orbital periods ranging from 10 to 1000 days, with velocity semi-amplitudes $K_1 \lesssim 50$~${\rm km\,s^{-1}}$ and mass functions $f(M_2)$ between 0.03 and 0.94~$M_{\odot}$. The selection process, as detailed in Figure~\ref{fig:flowchart}, relied on identifying long-term RV variations while ensuring stability over short time scales and excluding double-lined spectroscopic binaries.

For each candidate, we determined stellar parameters ($T_{\rm eff}$, $R_1$, $\log g$, and [Fe/H]) through SED fitting and estimated isochrone masses ($M_1$). Based on the derived $f(M_2)$, $M_1$, and $R_1$, the 74 candidates were classified into three categories: Class A includes five strong compact object candidates with main-sequence companions; Class B consists of 9 binaries with giant or subgiant companions; and Class C contains 61 systems with lower mass ratios. 

\textit{Gaia} DR3 \texttt{nss\_two\_star\_orbit} catalog provides 16 candidates with orbital solutions, all of which show consistency with our RV-based results. These findings demonstrate that long-term spectroscopic monitoring, particularly through surveys like LAMOST, is highly effective for uncovering the population of wide-orbit binaries and potential compact object hosts.

Further constraining the exact mass of the compact object remains challenging due to the unknown orbital inclination $i$. For these long-period systems, $v \sin i$ measurements cannot be reliably used to infer $i$ because tidal alignment between the stellar rotation axis and the orbital axis is not expected. Furthermore, in these long-period systems, the visible companions are far from filling their Roche lobes; consequently, the absence of significant ellipsoidal variations in their light curves precludes the use of photometric modeling to constrain the inclination. While targets with Thiele-Innes elements from \textit{Gaia} provide some initial constraints, the issue of orbital inclination is expected to be more robustly addressed as more epoch astrometry becomes available in future \textit{Gaia} data releases.

The candidate selection in this study exhibits a potential concentration toward late-type stars (spectral types F to M), a distribution that originates from fundamental physical and observational constraints rather than merely the properties of the parent catalog. A primary driver is the necessity for high-precision RV measurements to resolve the subtle modulations characteristic of long-period orbits, where velocity semi-amplitudes are typically small. Late-type stars possess relatively cool stellar atmospheres with a high density of narrow metallic absorption lines, which significantly enhance the signal-to-noise ratio of the CCF. In contrast, early-type stars often feature sparse spectra and significant line broadening due to high rotational velocities or thermal effects, making them less suitable for detecting low-amplitude RV variations. Consequently, these late-type stars serve as the most reliable dynamical tracers for identifying wide-separation, long-period systems.

Finally, the survival and detection of such systems are influenced by the formation history of the compact object. Middle-to-low mass stars, which comprise the majority of our late-type sample, have significantly longer main-sequence lifetimes compared to massive stars, providing a much broader temporal window for observation. Moreover, the formation of a compact object through a supernova event in a massive binary often involves a significant mass loss and a natal kick that can disrupt wide-orbit systems. Systems that remain bound with a late-type companion often represent a population that has survived these dynamical perturbations. Thus, the composition of our sample reflects a combination of the precision required for RV monitoring and the physical conditions necessary for the survival of wide-separation binaries in the Galaxy.

 \section*{acknowledgments}
We thank Tuan Yi and Xinlin Zhao for the helpful discussion, and thank the anonymous referee for the insightful suggestions.
This work was supported by the National Key R\&D Program of China under grants 2021YFA1600401 and 2023YFA1607901, the National Natural Science Foundation of China under grants 12433007 and 12221003. 
We acknowledge the science research grants from the China Manned Space Project with No. CMS-CSST-2025-A13.
This paper uses the data from the LAMOST survey.
Guoshoujing Telescope (the Large Sky Area Multi-Object Fiber Spectroscopic Telescope LAMOST) is a National Major Scientific Project built by the Chinese Academy of Sciences. 
Funding for the project has been provided by the National Development and Reform Commission. LAMOST is operated and managed by the National Astronomical Observatories, Chinese Academy of Sciences.

 \software{
    Astropy \citep{2013A&A...558A..33A,2018AJ....156..123A,2022ApJ...935..167A}, 
    Emcee \citep{foreman2013emcee}, 
    StellarSpecModel \citep{stellarSpecModel2025}, 
    Spectool \citep{spectool2025}, 
    ARIADNE \citep{Vines_Jenkins2022}, 
    Isochrones \citep{paxton2015}, MultiNest \citep{feroz2008,feroz2009,feroz2019}, PyMultinest \citep{buchner2014}
}

\bibliography{main}{}
\bibliographystyle{aasjournalv7}

\appendix
\section{Illustrative Orbital Solutions}
\label{sec:rv_solutions}
\begin{figure*}[h] 
    \centering
    \subfloat[J005516.76+061628.2]{\includegraphics[width=0.32\textwidth]{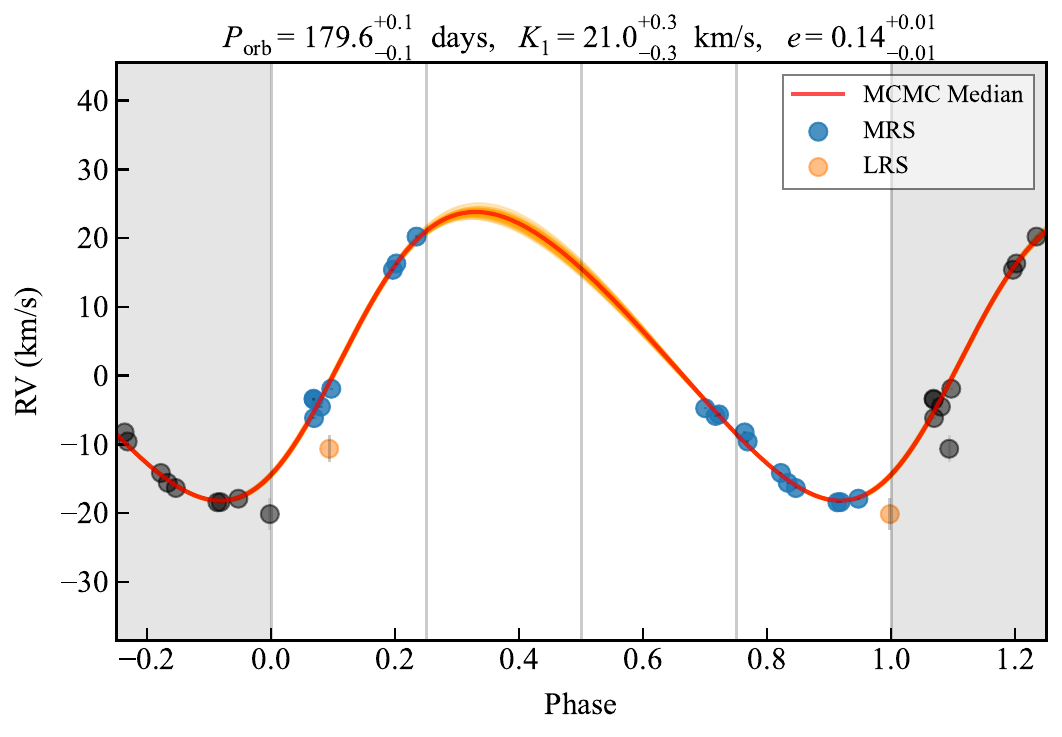}\label{fig:rv_ex1}}
    \hfill
    \subfloat[J061923.86+173856.6]{\includegraphics[width=0.32\textwidth]{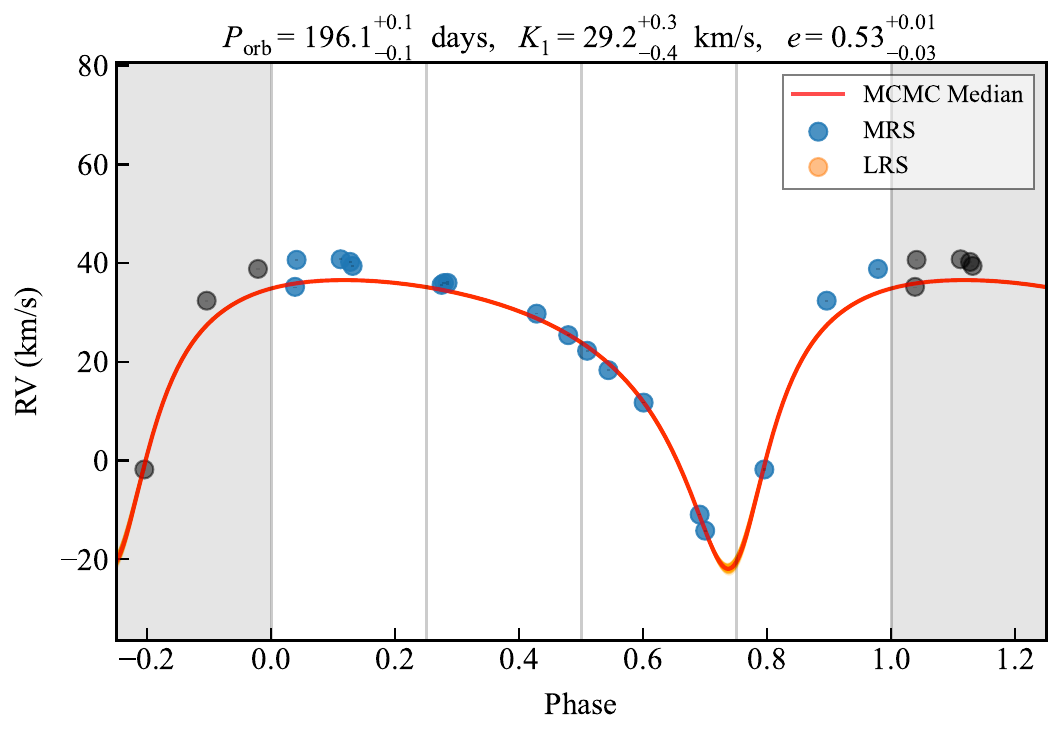}\label{fig:rv_ex2}}
    \hfill
    \subfloat[J075701.80+421519.3]{\includegraphics[width=0.32\textwidth]{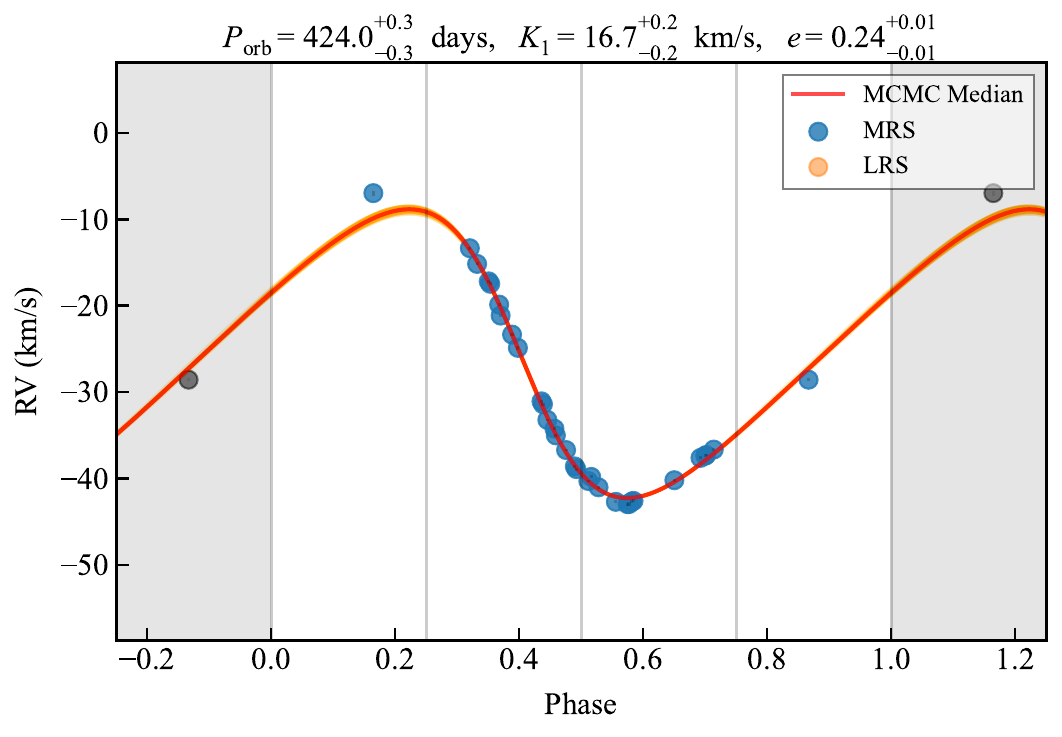}\label{fig:rv_ex3}}
    \\ 
    \subfloat[J064915.62+234352.8]{\includegraphics[width=0.32\textwidth]{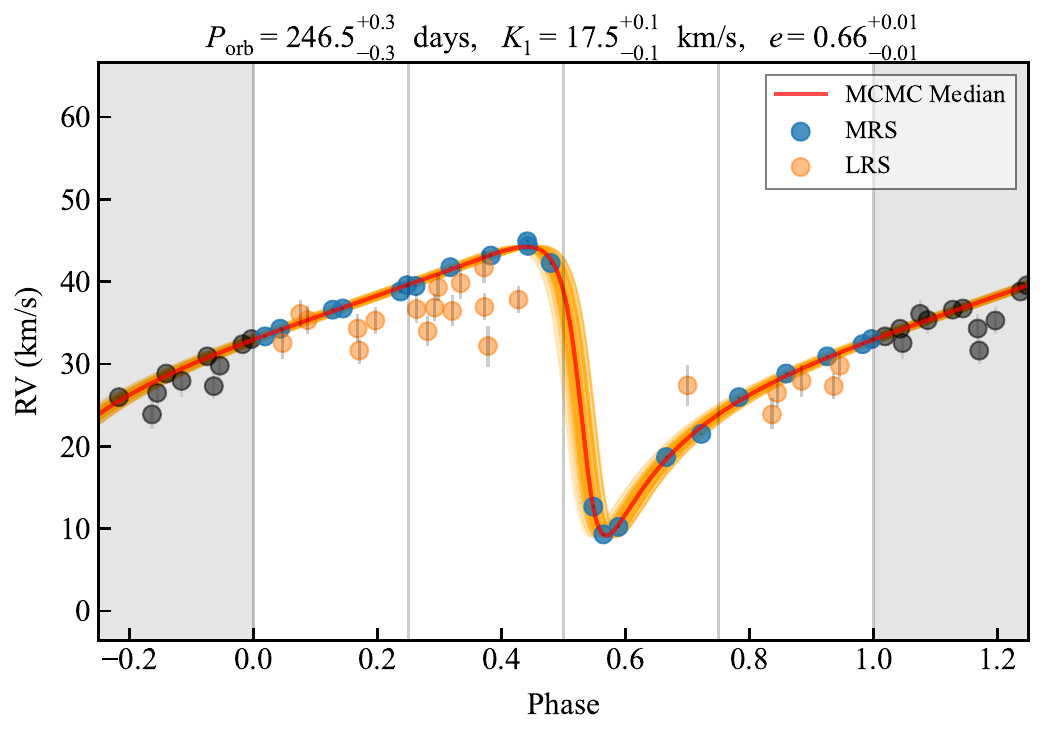}\label{fig:rv_ex4}}
    \hfill
    \subfloat[J030035.16+560301.9]{\includegraphics[width=0.32\textwidth]{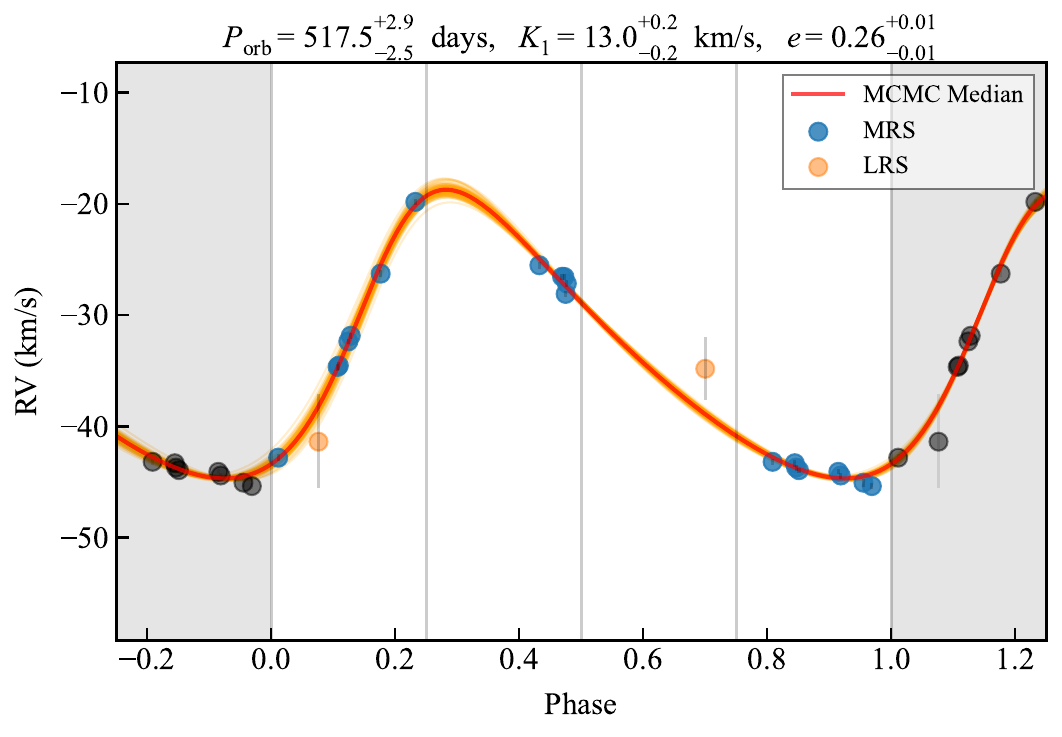}\label{fig:rv_ex5}}
    \hfill
    \subfloat[J030010.36+544051.5]{\includegraphics[width=0.32\textwidth]{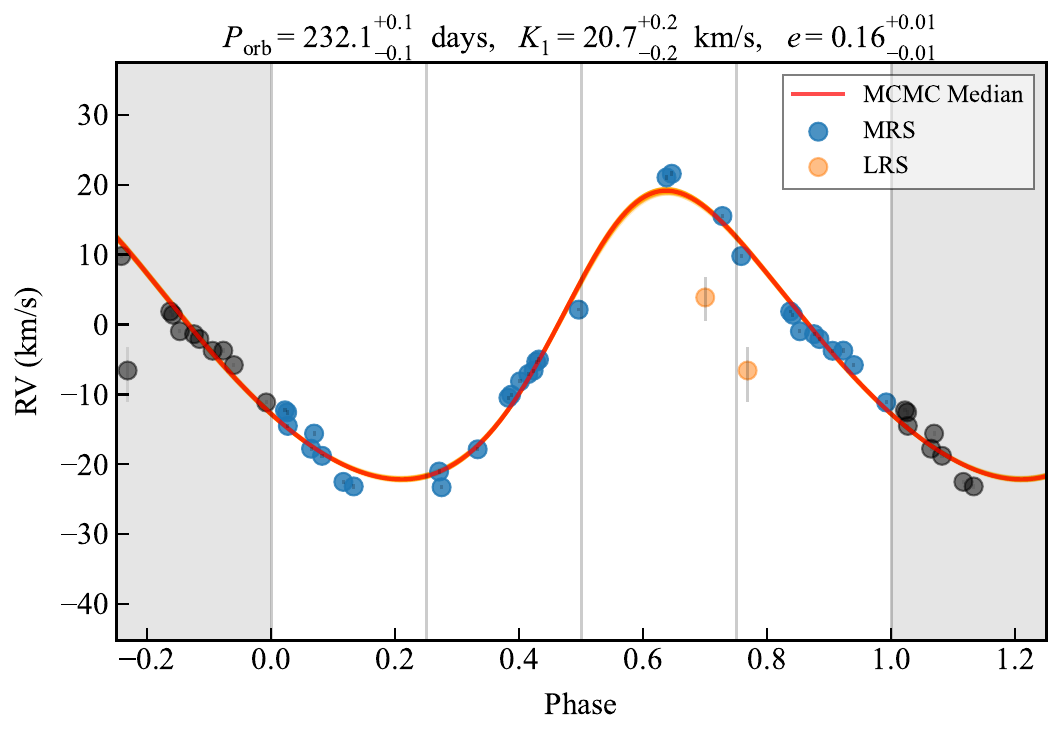}\label{fig:rv_ex6}}
    \\ 
    \subfloat[J044227.76+240747.0]{\includegraphics[width=0.32\textwidth]{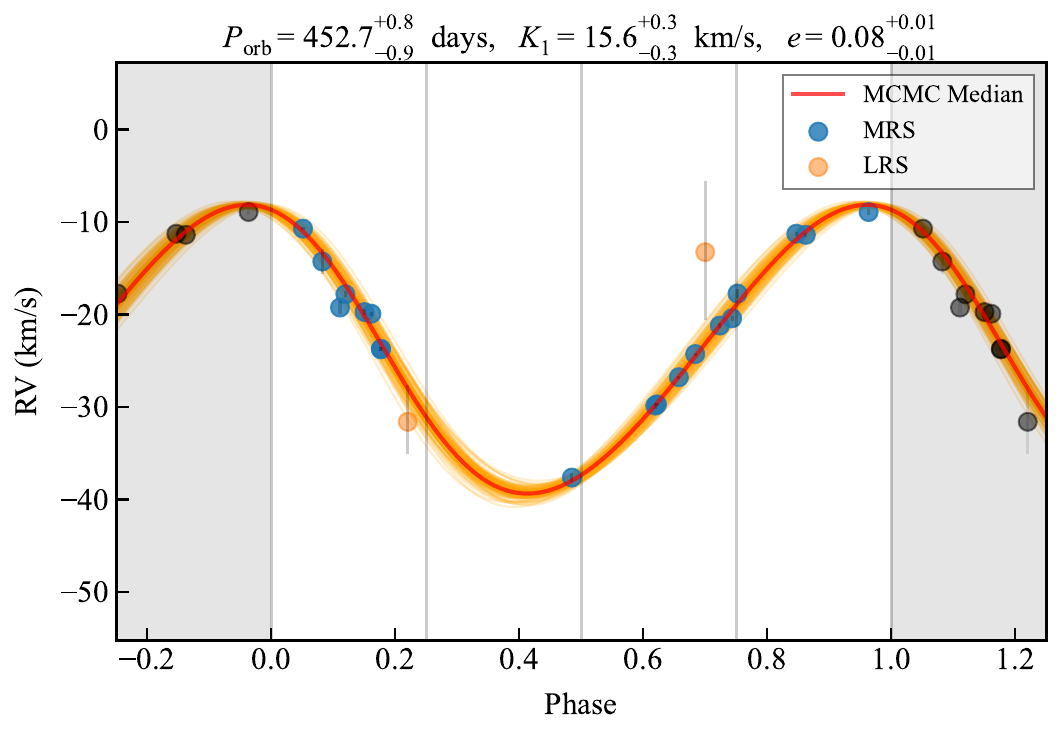}\label{fig:rv_ex7}}
    \hfill
    \subfloat[J044741.50+483453.0]{\includegraphics[width=0.32\textwidth]{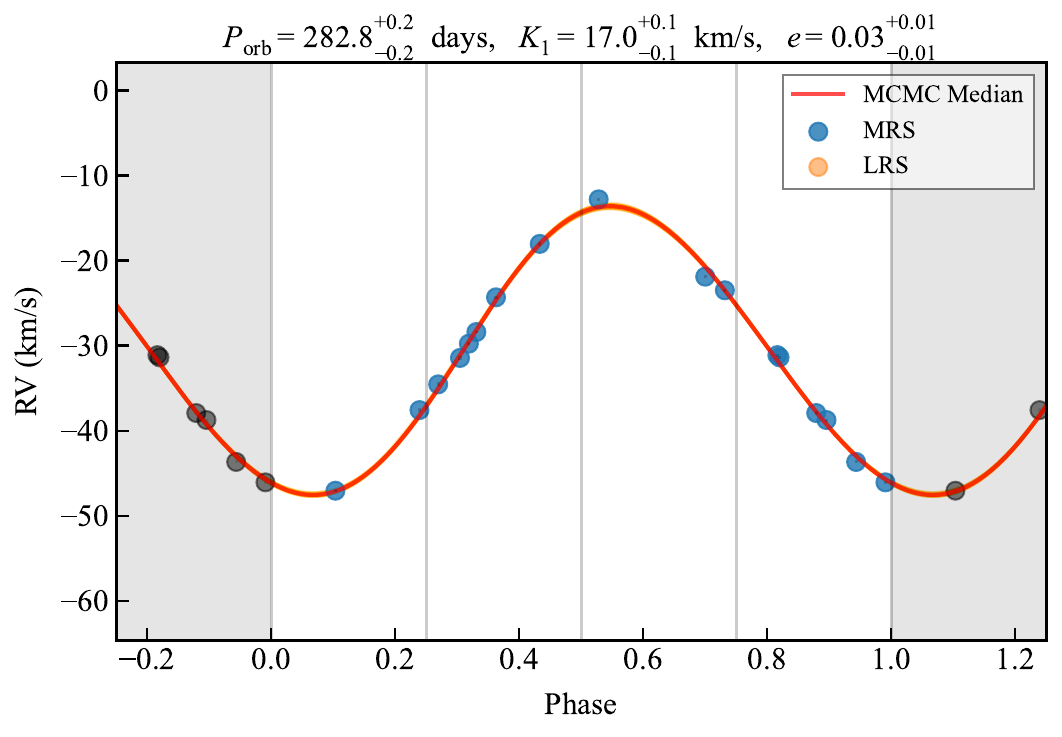}\label{fig:rv_ex8}}
    \hfill
    \subfloat[J050447.62+482902.9]{\includegraphics[width=0.32\textwidth]{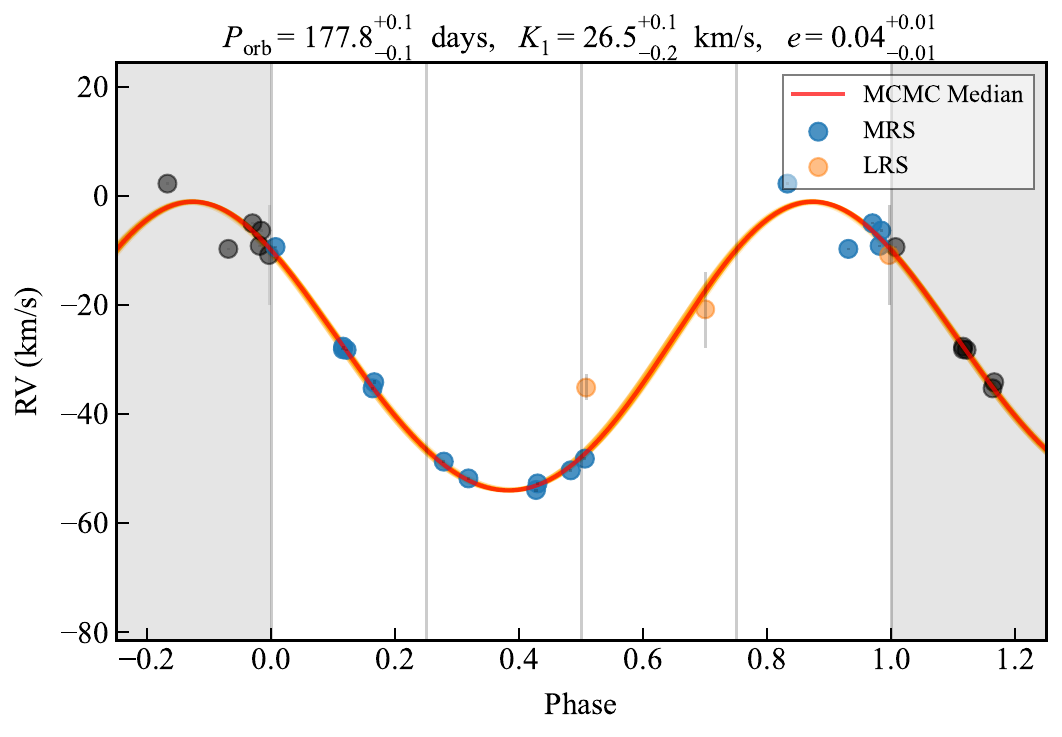}\label{fig:rv_ex9}}
    \caption{Phase-folded RV curves and best-fit Keplerian orbital solutions for nine representative binary candidates from our sample. For each panel, blue circles denote the MRS observations, orange circles denote the LRS observations. The solid orange line indicates the median fit, while the shaded region represents the bundle of possible orbits consistent with the posterior distribution.}
    \label{fig:rv_examples}
\end{figure*}

In this section, we present a selection of RV orbital solutions for nine representative candidates from our sample. Each panel in Figure \ref{fig:rv_examples} displays the phase-folded RV curve for a single candidate, along with its best-fit Keplerian orbital solution. To derive the orbital solutions, we employed a MCMC sampler to fit the key parameters, including $P_{\text{orb}}$, e and $K_1$.
The detailed orbital parameters for these and all other candidates are comprehensively listed in Tables \ref{tab:orbit} in Appendix \ref{sec:full_candidates}.

\clearpage
\begin{figure*}[b]
    \centering
    \subfloat[Corner plot for J024859.02+534307.2 ]{
        \includegraphics[width=0.48\textwidth]{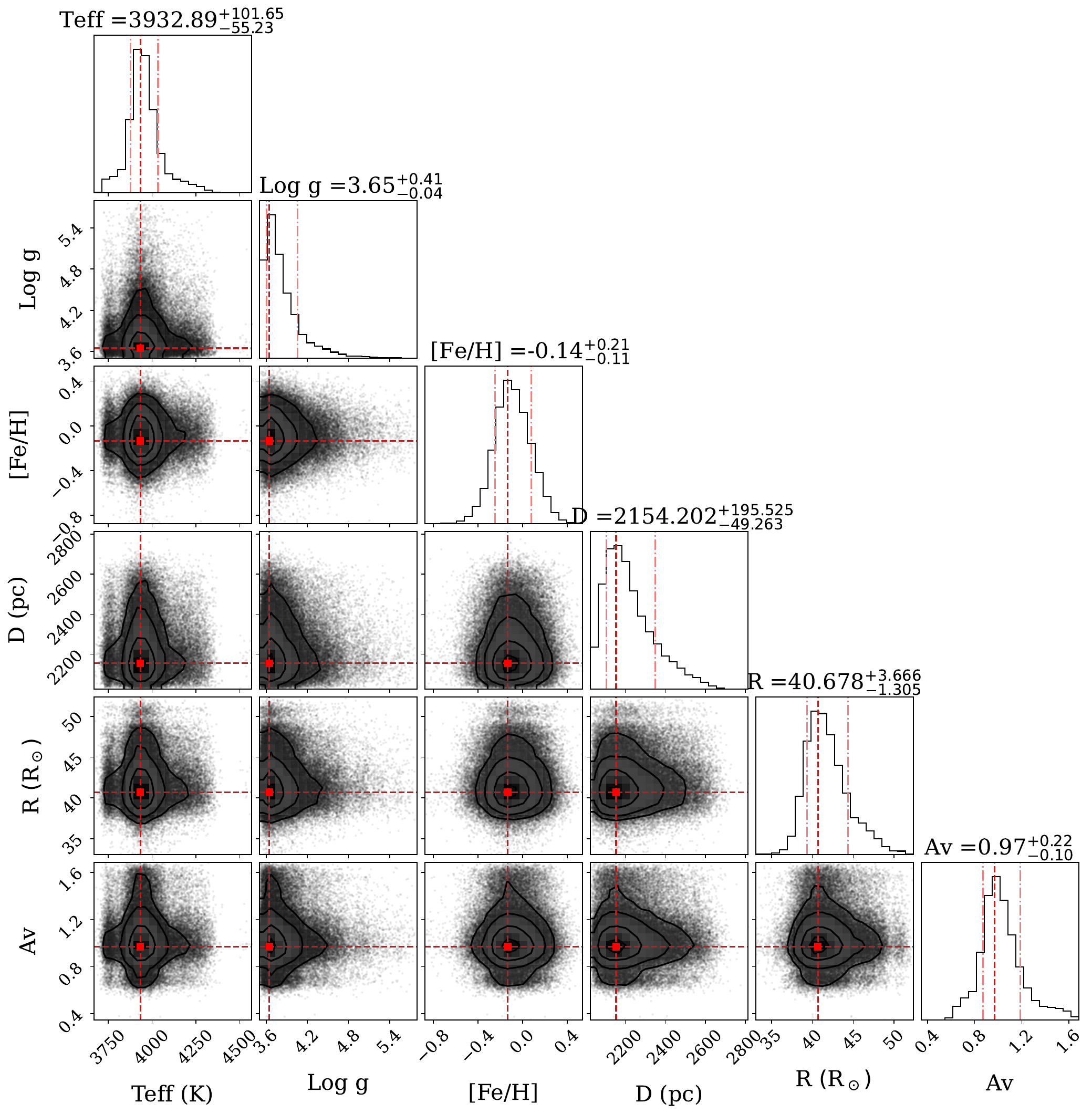}
    }
    \hfill
    \subfloat[Corner plot for J085413.00+125553.1]{
        \includegraphics[width=0.48\textwidth]{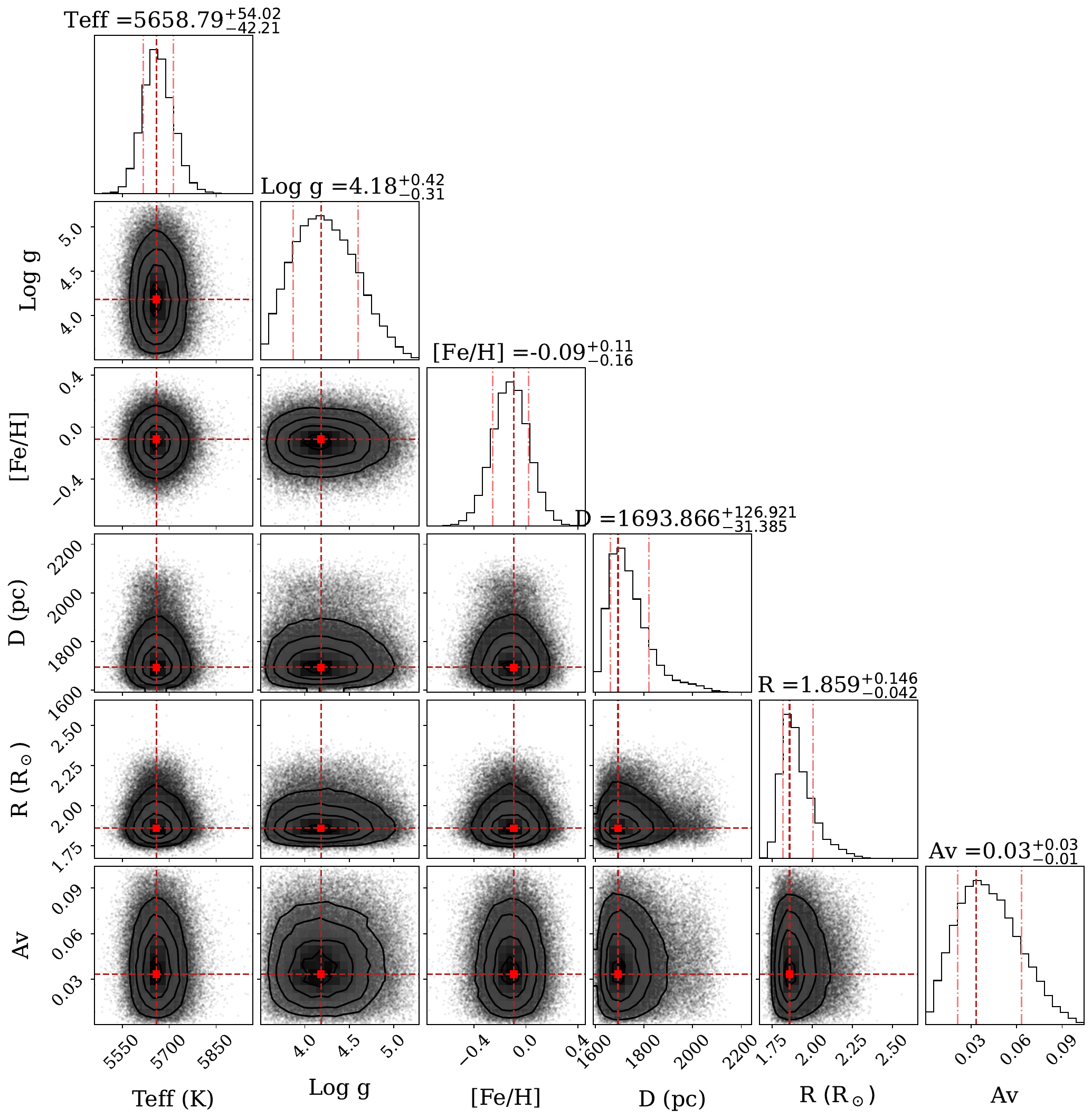}
    }

    \vspace{0.5cm}

    \subfloat[Best-fit SED for J024859.02+534307.2 ]{
        \includegraphics[width=0.48\textwidth]{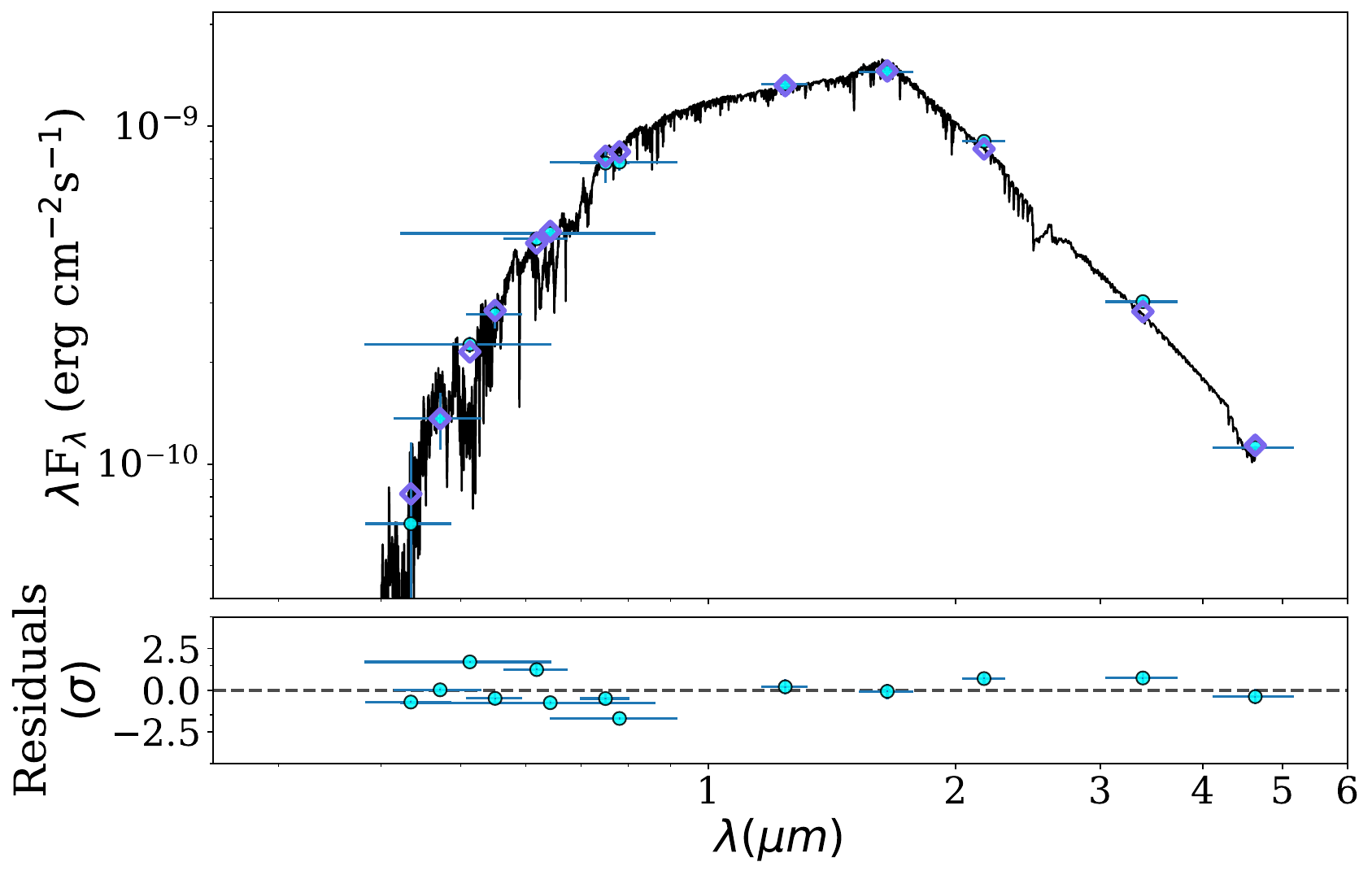}
    }
    \hfill
    % one
    \subfloat[Best-fit SED for J085413.00+125553.1]{
        \includegraphics[width=0.48\textwidth]{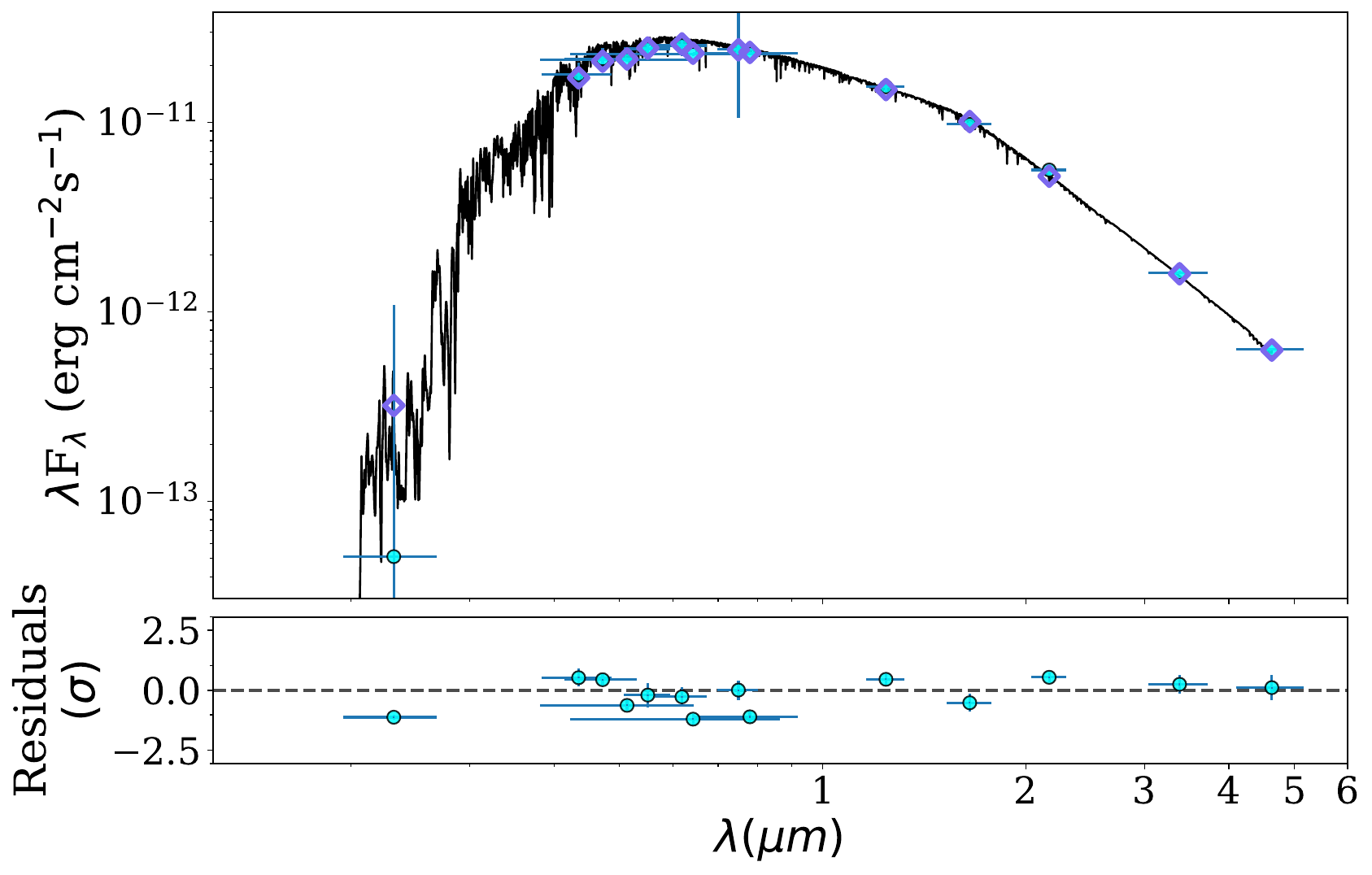}
    }

    \caption{Representative SED fitting results and posterior probability distributions (corner plots) for two candidates. In the SED panels, the black line represents the best-fit model, and the colored points denote the observed photometry. The corner plots show the correlations between the derived stellar parameters ($T_{\rm eff}$, $R_1$, [Fe/H], and $\log g$). }
    \label{fig:sed_appendix}

\end{figure*}

\section{SED Fitting}
\label{app:sed_details}

This section details the technical nuances of our SED modeling and the stringent criteria for photometric data selection. To ensure the fidelity of the derived stellar parameters, we implemented a targeted data-filtering strategy. While \textit{astroARIADNE} can incorporate a wide array of catalogs, we found that certain datasets (most notably SDSS and TESS) frequently introduced systematic biases for our bright candidate stars possibly due to saturation effects and pixel-level blending. Consequently, these were excluded to mitigate noise in the Bayesian model-averaging process.

Instead, our analysis prioritized a core set of high-reliability, multi-band catalogs to ensure broad panchromatic coverage. We utilized the Galaxy Evolution Explorer (GALEX) for ultraviolet constraints, the AAVSO Photometric All-Sky Survey (APASS) for the optical regime, and the Two Micron All-Sky Survey (2MASS) coupled with the Wide-field Infrared Survey Explorer (WISE) for the near- and mid-infrared. The resulting SED fits and the associated posterior probability distributions (corner plots) are illustrated in Figure \ref{fig:sed_appendix}, where the highly localized constraints on $R_1$ and $T_{\rm eff}$ underscore the robustness of our filtered-photometry approach.

\section{Summary of Derived Parameters for Other Candidates}
\label{sec:full_candidates}
In this appendix, we summarize the derived parameters for the remaining candidates in our sample. The results of RV measurements are listed in Table \ref{tab:rv}, where only a representative sample is shown and the complete table is available online. The orbital solutions derived from these measurements are presented in Table \ref{tab:orbit}. Finally, the stellar parameters obtained from the SED fitting are provided in Table \ref{tab:params}.

\begin{longtable}{@{\extracolsep{\fill}}clcccccccc}
\caption{Summary of RV Measurements.}\\
\toprule
\hline
    SID & Source & \texttt{obsid} & \texttt{lmjd} & \texttt{date} & \texttt{lmjm} & \texttt{snr} &    RV & Uncertainty & Residual \\
    &  &  &  &  &  &  & (${\rm km\,s^{-1}}$) & (${\rm km\,s^{-1}}$) & (${\rm km\,s^{-1}}$) \\
(1) & (2) & (3) & (4) & (5) & (6) & (7) & (8) & (9) & (10)\\
\midrule
\endfirsthead
\caption[]{Summary of RV Measurements.} \\
\toprule
    SID & Source & \texttt{obsid} & \texttt{lmjd} & \texttt{date} & \texttt{lmjm} & \texttt{snr} &    RV & Uncertainty & Residual \\
    &  &  &  &  &  &  & (${\rm km\,s^{-1}}$) & (${\rm km\,s^{-1}}$) & (${\rm km\,s^{-1}}$) \\
(1) & (2) & (3) & (4) & (5) & (6) & (7) & (8) & (9) & (10)\\
\midrule
\endhead
\midrule
\multicolumn{10}{r}{{Continued on next page}} \\
\midrule
\endfoot

\bottomrule
\endlastfoot
1098054 &    LRS &      127806239 &         56350 &    2013-02-26 &      81143772 &       269.88 & 57.79 &        1.92 &    -1.83 \\
        &    LRS &      212106239 &         56684 &    2014-01-26 &      81624849 &       136.06 & 28.82 &        2.34 &     1.63 \\
        &  MRS-B &      597006239 &         58060 &    2017-11-02 &      83606654 &         6.78 & 80.18 &        0.97 &    -1.29 \\
        & MRS-BR &      726806239 &         58532 &    2019-02-17 &      84285881 &       290.10 & 68.87 &        0.14 &    -1.29 \\
        & MRS-BR &      727406239 &         58536 &    2019-02-21 &      84291625 &       350.07 & 74.81 &        0.13 &     0.23 \\
        & MRS-BR &      766506233 &         58806 &    2019-11-18 &      84680774 &       272.67 & 92.24 &        0.14 &     1.25 \\
        &  MRS-B &      771906233 &         58823 &    2019-12-05 &      84705213 &       200.91 & 82.73 &        0.16 &    -0.56 \\
        &    ... &            ... &           ... &           ... &           ... &          ... &   ... &         ... &      ... \\
\hline
1753900 &    LRS &      420916159 &         57425 &    2016-02-06 &      82691961 &       185.81 & 45.64 &        7.06 &    -2.34 \\
        &  MRS-B &      607116159 &         58084 &    2017-11-26 &      83641262 &        68.42 & 62.14 &        0.31 &    -1.53 \\
        &  MRS-B &      607816159 &         58086 &    2017-11-28 &      83644134 &        70.34 & 62.70 &        0.25 &    -1.01 \\
        &  MRS-B &      608716159 &         58088 &    2017-11-30 &      83647007 &        32.63 & 61.50 &        0.27 &    -2.25 \\
        &  MRS-B &      634916159 &         58144 &    2018-01-25 &      83727399 &        69.73 & 61.56 &        0.24 &    -2.84 \\
        & MRS-BR &      640816159 &         58153 &    2018-02-03 &      83740363 &        92.67 & 62.25 &        0.37 &    -2.18 \\
        &  MRS-B &      653416159 &         58182 &    2018-03-04 &      83782026 &        52.68 & 61.00 &        0.24 &    -3.40 \\
        &    ... &            ... &           ... &           ... &           ... &          ... &   ... &         ... &      ... \\
\hline
1767926 &    LRS &      339814061 &         57119 &    2015-04-06 &      82251145 &        75.22 & 14.15 &        2.84 &    -2.15 \\
        &    LRS &      432713076 &         57444 &    2016-02-25 &      82719321 &        45.52 & -6.92 &        2.94 &    -0.18 \\
        & MRS-BR &      695802019 &         58452 &    2018-11-29 &      84171124 &         4.18 & 10.88 &        1.21 &    -0.87 \\
        & MRS-BR &      703002019 &         58470 &    2018-12-17 &      84197024 &        82.96 & 16.47 &        0.15 &     0.51 \\
        & MRS-BR &      717802019 &         58504 &    2019-01-20 &      84245861 &        19.17 & 23.96 &        0.29 &     1.09 \\
        & MRS-BR &      719302019 &         58508 &    2019-01-24 &      84251607 &        40.42 & 23.41 &        0.15 &    -0.06 \\
        & MRS-BR &      719902019 &         58509 &    2019-01-25 &      84253032 &         8.15 & 23.03 &        0.61 &    -0.57 \\
        &    ... &            ... &           ... &           ... &           ... &          ... &   ... &         ... &      ... \\
\label{tab:rv}
\end{longtable}

\begin{minipage}{\textwidth}
    \vspace{-2pt} 
    \footnotesize
    \textbf{Notes.} \\
    Column (1): Source ID in our sample; 
    Column (2): Spectroscopic survey of the data; 
    Columns (3) - (7): The \texttt{obsid}, \texttt{lmjd}, \texttt{date}, \texttt{lmjm} and \texttt{snr} fields of LAMOST catalog; 
    Column (8): Radial velocity; 
    Column (9): Radial velocity uncertainty; 
    Column (10): Residuals to orbit listed in Table \ref{tab:orbit}. \\
$\ast$ Table \ref{tab:rv} is published in its entirety in the machine-readable format. A portion is shown here for guidance regarding its form and content.
\end{minipage}

\digitalasset{}

\clearpage

\begin{longtable}{@{\extracolsep{-1pt}}crrrrrrrrc}
\caption{Orbital Solutions for the Candidates}\\
\toprule
\hline
LAMOST designation & SID & $P_{\rm orb}$ & $K_1$ & e & $\gamma$ & $\omega$ & $t_0$ & $f(M_2)$ & Class \\
& & (days) & (${\rm km\,s^{-1}}$) &  & (${\rm km\,s^{-1}}$) &  & (\texttt{lmjd}) & ($M_{\odot}$) & \\
(1) & (2) & (3) & (4) & (5) & (6) & (7) & (8) & (9) & (10)\\
\midrule
\endfirsthead
\caption[]{Orbital Solutions — Continued} \\
\toprule
LAMOST designation & SID & $P_{\rm orb}$ & $K_1$ & e & $\gamma$ & $\omega$ & $t_0$ & $f(M_2)$ & Class \\
& & (days) & (${\rm km\,s^{-1}}$) &  & (${\rm km\,s^{-1}}$) &  & (\texttt{lmjd}) & ($M_{\odot}$) & \\
(1) & (2) & (3) & (4) & (5) & (6) & (7) & (8) & (9) & (10)\\
\midrule
\endhead
\midrule
\multicolumn{9}{r}{{Continued on next page}} \\
\midrule
\endfoot

\bottomrule
\endlastfoot
J060228.55+280758.8 & 1098054 &   $120.8^{+0.1}_{-0.2}$ & $31.9^{+0.1}_{-1.1}$ & $0.20^{+0.01}_{-0.12}$ &   $64.3^{+1.1}_{-0.1}$ &  $2.56^{+0.15}_{-0.02}$ &    $59575.6^{+2.1}_{-0.3}$ & $0.38^{+0.02}_{-0.04}$ &     A \\
J084314.39+133049.6 & 1753900 &  $1127.9^{+3.6}_{-3.3}$ & $11.5^{+0.2}_{-0.1}$ & $0.09^{+0.01}_{-0.01}$ &   $54.0^{+0.1}_{-0.1}$ &  $3.12^{+0.02}_{-0.04}$ &    $59850.2^{+4.0}_{-5.9}$ & $0.17^{+0.01}_{-0.01}$ &     A \\
J084423.83+152520.8 & 1767926 &  $450.9^{+3.3}_{-23.6}$ & $17.3^{+0.4}_{-0.7}$ & $0.15^{+0.24}_{-0.04}$ &    $5.6^{+0.6}_{-2.5}$ &  $0.34^{+0.23}_{-0.97}$ & $59454.3^{+19.0}_{-121.5}$ & $0.21^{+0.03}_{-0.04}$ &     A \\
J085354.59+112346.6 & 1795960 &   $236.1^{+0.1}_{-0.1}$ & $30.3^{+0.3}_{-0.4}$ & $0.61^{+0.01}_{-0.01}$ &   $32.5^{+0.2}_{-0.1}$ &  $0.81^{+0.02}_{-0.02}$ &    $59055.9^{+0.2}_{-0.3}$ & $0.33^{+0.02}_{-0.02}$ &     A \\
J025948.47+544607.7 &  441454 &   $325.3^{+0.2}_{-0.2}$ & $28.5^{+0.2}_{-0.2}$ & $0.55^{+0.01}_{-0.01}$ &  $-56.1^{+0.2}_{-0.2}$ &  $0.48^{+0.02}_{-0.02}$ &    $59144.8^{+0.3}_{-0.3}$ & $0.45^{+0.01}_{-0.01}$ &     B \\
J030010.36+544051.5 &  454174 &   $232.1^{+0.1}_{-0.1}$ & $20.7^{+0.2}_{-0.2}$ & $0.16^{+0.01}_{-0.01}$ &   $-3.9^{+0.1}_{-0.1}$ & $-0.78^{+0.03}_{-0.03}$ &    $58851.5^{+1.1}_{-1.0}$ & $0.20^{+0.01}_{-0.01}$ &     B \\
J030058.13+533839.1 &  454319 &    $37.6^{+0.1}_{-0.1}$ & $36.7^{+0.1}_{-0.2}$ & $0.01^{+0.01}_{-0.01}$ &  $-13.8^{+0.1}_{-0.1}$ & $-1.85^{+0.30}_{-0.20}$ &    $59865.1^{+1.8}_{-1.2}$ & $0.19^{+0.01}_{-0.01}$ &     B \\
J031324.69+545919.5 &  488064 &   $274.0^{+0.2}_{-0.2}$ & $32.5^{+0.2}_{-0.2}$ & $0.09^{+0.01}_{-0.01}$ &  $-33.9^{+0.2}_{-0.2}$ & $-1.66^{+0.01}_{-0.01}$ &    $58675.3^{+0.2}_{-0.1}$ & $0.94^{+0.02}_{-0.02}$ &     B \\
J043457.76+275049.4 &  782766 &   $577.4^{+0.9}_{-0.8}$ & $17.1^{+0.2}_{-0.2}$ & $0.39^{+0.01}_{-0.01}$ &   $-9.7^{+0.2}_{-0.2}$ &  $2.76^{+0.03}_{-0.03}$ &    $59863.2^{+1.6}_{-1.8}$ & $0.23^{+0.01}_{-0.01}$ &     B \\
J045853.08+482528.9 &  865354 &    $31.2^{+0.1}_{-0.1}$ & $44.8^{+0.1}_{-0.1}$ & $0.00^{+0.01}_{-0.01}$ &  $-18.5^{+0.1}_{-0.1}$ & $-0.37^{+0.35}_{-0.12}$ &    $59495.2^{+1.7}_{-0.6}$ & $0.28^{+0.01}_{-0.01}$ &     B \\
J050207.99+500019.9 &  877473 & $1077.8^{+37.2}_{-7.6}$ & $18.0^{+0.4}_{-0.3}$ & $0.41^{+0.02}_{-0.03}$ &  $-16.2^{+0.3}_{-0.4}$ & $-3.04^{+0.04}_{-0.04}$ &   $58532.4^{+8.7}_{-26.9}$ & $0.49^{+0.04}_{-0.03}$ &     B \\
J050259.65+454413.1 &  878211 &    $41.1^{+0.1}_{-0.1}$ & $45.9^{+0.2}_{-0.2}$ & $0.00^{+0.01}_{-0.01}$ &   $-6.8^{+0.1}_{-0.1}$ &  $1.88^{+0.64}_{-1.06}$ &    $59198.7^{+4.2}_{-6.9}$ & $0.40^{+0.01}_{-0.01}$ &     B \\
J091645.31+431251.8 & 1890750 &   $152.0^{+0.1}_{-0.1}$ & $23.4^{+0.1}_{-0.1}$ & $0.06^{+0.01}_{-0.01}$ &   $-9.5^{+0.1}_{-0.1}$ &  $1.79^{+0.05}_{-0.05}$ &    $58953.6^{+1.2}_{-1.2}$ & $0.20^{+0.01}_{-0.01}$ &     B \\
J005516.76+061628.2 &  109269 &   $179.6^{+0.1}_{-0.1}$ & $21.0^{+0.3}_{-0.3}$ & $0.14^{+0.01}_{-0.01}$ &    $3.5^{+0.3}_{-0.2}$ & $-1.80^{+0.03}_{-0.03}$ &    $59568.3^{+0.7}_{-0.7}$ & $0.16^{+0.01}_{-0.01}$ &     C \\
J024859.02+534307.2 &  422741 &   $458.9^{+1.0}_{-1.1}$ & $14.4^{+0.1}_{-0.1}$ & $0.27^{+0.01}_{-0.01}$ &   $32.3^{+0.1}_{-0.1}$ &  $2.03^{+0.02}_{-0.03}$ &    $58906.3^{+1.5}_{-1.5}$ & $0.12^{+0.01}_{-0.01}$ &     C \\
J030035.16+560301.8 &  453063 &   $517.5^{+2.9}_{-2.5}$ & $13.0^{+0.2}_{-0.2}$ & $0.26^{+0.01}_{-0.01}$ &  $-33.3^{+0.2}_{-0.2}$ & $-1.08^{+0.07}_{-0.06}$ &    $59704.3^{+6.6}_{-5.4}$ & $0.10^{+0.01}_{-0.01}$ &     C \\
J034116.64+492438.1 &  576683 &   $488.2^{+0.4}_{-0.4}$ & $21.1^{+0.2}_{-0.2}$ & $0.11^{+0.01}_{-0.01}$ &  $-26.7^{+0.1}_{-0.1}$ & $-1.42^{+0.07}_{-0.07}$ &    $59163.0^{+4.5}_{-4.6}$ & $0.45^{+0.01}_{-0.01}$ &     C \\
J035243.07+251621.9 &  623589 &   $224.4^{+0.1}_{-0.1}$ & $15.6^{+0.1}_{-0.1}$ & $0.23^{+0.01}_{-0.01}$ &  $-12.4^{+0.1}_{-0.1}$ & $-0.93^{+0.03}_{-0.02}$ &    $58189.6^{+0.9}_{-0.6}$ & $0.08^{+0.01}_{-0.01}$ &     C \\
J041014.93+533012.6 &  686208 &   $158.1^{+0.1}_{-0.1}$ & $24.4^{+0.3}_{-0.3}$ & $0.21^{+0.01}_{-0.01}$ &  $-34.9^{+0.4}_{-0.4}$ & $-0.44^{+0.08}_{-0.07}$ &    $58821.1^{+1.6}_{-1.4}$ & $0.22^{+0.01}_{-0.01}$ &     C \\
J044227.76+240747.1 &  808439 &   $452.7^{+0.8}_{-0.9}$ & $15.6^{+0.3}_{-0.3}$ & $0.08^{+0.01}_{-0.01}$ &  $-24.0^{+0.2}_{-0.2}$ &  $1.34^{+0.16}_{-0.17}$ &  $59205.2^{+10.5}_{-11.4}$ & $0.17^{+0.01}_{-0.01}$ &     C \\
J044741.50+483453.0 &  824490 &   $282.8^{+0.2}_{-0.2}$ & $17.0^{+0.1}_{-0.1}$ & $0.03^{+0.01}_{-0.01}$ &  $-30.7^{+0.1}_{-0.1}$ & $-1.34^{+0.13}_{-0.15}$ &    $59539.7^{+5.8}_{-6.4}$ & $0.14^{+0.01}_{-0.01}$ &     C \\
J050447.62+482902.8 &  888996 &   $177.8^{+0.1}_{-0.1}$ & $26.5^{+0.1}_{-0.2}$ & $0.04^{+0.01}_{-0.01}$ &  $-28.6^{+0.1}_{-0.1}$ & $-0.36^{+0.02}_{-0.01}$ &    $58772.4^{+0.4}_{-0.2}$ & $0.33^{+0.01}_{-0.01}$ &     C \\
J050636.51+483830.8 &  889042 &   $227.3^{+0.1}_{-0.1}$ & $28.5^{+0.2}_{-0.2}$ & $0.20^{+0.01}_{-0.01}$ &  $-25.2^{+0.1}_{-0.1}$ &  $1.79^{+0.02}_{-0.02}$ &    $59248.8^{+0.6}_{-0.7}$ & $0.50^{+0.01}_{-0.01}$ &     C \\
J054945.82+284136.3 & 1047481 &   $442.9^{+0.3}_{-0.3}$ & $13.2^{+0.1}_{-0.1}$ & $0.34^{+0.01}_{-0.01}$ &   $-9.2^{+0.1}_{-0.1}$ & $-1.59^{+0.01}_{-0.01}$ &    $59765.5^{+0.2}_{-0.1}$ & $0.09^{+0.01}_{-0.01}$ &     C \\
J055223.94+282946.1 & 1063997 &    $56.4^{+0.1}_{-0.1}$ & $36.2^{+0.1}_{-0.1}$ & $0.01^{+0.01}_{-0.01}$ &   $19.6^{+0.1}_{-0.1}$ & $-0.61^{+0.16}_{-0.18}$ &    $59942.6^{+1.5}_{-1.6}$ & $0.27^{+0.01}_{-0.01}$ &     C \\
J055726.63+293812.6 & 1080054 &   $197.0^{+0.1}_{-0.1}$ & $24.7^{+0.2}_{-0.2}$ & $0.40^{+0.01}_{-0.01}$ &   $-6.3^{+0.1}_{-0.1}$ &  $0.01^{+0.01}_{-0.01}$ &    $59585.1^{+0.3}_{-0.3}$ & $0.23^{+0.01}_{-0.01}$ &     C \\
J061923.86+173856.7 & 1167929 &   $196.1^{+0.1}_{-0.1}$ & $29.2^{+0.3}_{-0.4}$ & $0.53^{+0.01}_{-0.03}$ &   $22.0^{+0.1}_{-0.1}$ & $-2.83^{+0.01}_{-0.01}$ &    $58425.0^{+0.2}_{-0.3}$ & $0.31^{+0.02}_{-0.02}$ &     C \\
J061815.50+191501.5 & 1171404 &   $152.0^{+0.1}_{-0.1}$ & $26.3^{+0.2}_{-0.2}$ & $0.09^{+0.01}_{-0.01}$ &   $-0.7^{+0.1}_{-0.1}$ &  $3.01^{+0.05}_{-0.06}$ &    $58505.5^{+1.2}_{-1.3}$ & $0.28^{+0.01}_{-0.01}$ &     C \\
J062027.21+170223.9 & 1186400 &   $406.1^{+0.4}_{-0.4}$ & $15.0^{+0.2}_{-0.2}$ & $0.33^{+0.01}_{-0.01}$ &   $27.4^{+0.2}_{-0.1}$ & $-0.70^{+0.02}_{-0.02}$ &    $58834.6^{+1.1}_{-1.0}$ & $0.12^{+0.01}_{-0.01}$ &     C \\
J062749.84+164036.9 & 1206050 &   $305.3^{+0.3}_{-0.3}$ & $19.8^{+0.1}_{-0.1}$ & $0.29^{+0.01}_{-0.01}$ &   $54.2^{+0.2}_{-0.2}$ & $-0.86^{+0.03}_{-0.02}$ &    $59861.9^{+1.1}_{-1.0}$ & $0.21^{+0.01}_{-0.01}$ &     C \\
J063343.31+185027.5 & 1243191 &    $38.7^{+0.4}_{-0.3}$ & $38.0^{+0.6}_{-0.5}$ & $0.04^{+0.04}_{-0.03}$ &   $23.7^{+0.5}_{-0.5}$ &  $0.09^{+1.02}_{-1.21}$ &  $59212.0^{+18.4}_{-13.0}$ & $0.21^{+0.01}_{-0.01}$ &     C \\
J064430.01+222242.3 & 1299572 &   $476.2^{+3.0}_{-3.8}$ & $12.6^{+0.3}_{-0.3}$ & $0.46^{+0.03}_{-0.05}$ &   $29.8^{+0.8}_{-0.6}$ & $-0.15^{+0.06}_{-0.05}$ &    $58660.0^{+2.8}_{-1.1}$ & $0.07^{+0.01}_{-0.01}$ &     C \\
J064915.62+234352.8 & 1316116 &   $246.5^{+0.3}_{-0.3}$ & $17.5^{+0.1}_{-0.1}$ & $0.66^{+0.01}_{-0.01}$ &   $31.5^{+0.1}_{-0.1}$ &  $2.00^{+0.01}_{-0.01}$ &    $58957.4^{+0.3}_{-0.3}$ & $0.06^{+0.01}_{-0.01}$ &     C \\
J075559.73+423618.3 & 1592874 &    $30.3^{+0.1}_{-0.1}$ & $24.3^{+0.2}_{-0.1}$ & $0.07^{+0.01}_{-0.01}$ &   $36.4^{+0.1}_{-0.1}$ &  $1.27^{+0.01}_{-0.01}$ &    $59990.1^{+0.1}_{-0.1}$ & $0.04^{+0.01}_{-0.01}$ &     C \\
J075753.54+400910.1 & 1606805 &    $36.0^{+0.1}_{-0.1}$ & $25.9^{+0.2}_{-0.2}$ & $0.02^{+0.01}_{-0.01}$ &   $74.1^{+0.2}_{-0.2}$ & $-1.12^{+0.34}_{-0.26}$ &    $58848.0^{+2.0}_{-1.6}$ & $0.06^{+0.01}_{-0.01}$ &     C \\
J075701.80+421519.6 & 1607121 &   $424.0^{+0.3}_{-0.3}$ & $16.7^{+0.2}_{-0.2}$ & $0.24^{+0.01}_{-0.01}$ &  $-25.3^{+0.1}_{-0.1}$ &  $1.63^{+0.02}_{-0.02}$ &    $59597.8^{+1.1}_{-1.2}$ & $0.18^{+0.01}_{-0.01}$ &     C \\
J080732.28+423044.7 & 1635668 &   $101.9^{+0.1}_{-0.1}$ & $21.9^{+0.1}_{-0.2}$ & $0.36^{+0.01}_{-0.01}$ &  $-51.3^{+0.1}_{-0.1}$ & $-0.39^{+0.01}_{-0.01}$ &    $59849.1^{+0.1}_{-0.1}$ & $0.09^{+0.01}_{-0.01}$ &     C \\
J080858.99+401829.2 & 1651164 &    $97.8^{+0.1}_{-0.1}$ & $18.4^{+0.2}_{-0.3}$ & $0.28^{+0.01}_{-0.02}$ &   $43.1^{+0.1}_{-0.1}$ & $-3.14^{+0.01}_{-0.01}$ &    $59279.7^{+0.2}_{-0.2}$ & $0.05^{+0.01}_{-0.01}$ &     C \\
J083758.91+120828.9 & 1738351 &    $75.2^{+0.1}_{-0.1}$ & $27.5^{+0.2}_{-0.2}$ & $0.53^{+0.01}_{-0.01}$ &   $26.6^{+0.1}_{-0.1}$ &  $0.99^{+0.01}_{-0.01}$ &    $59606.1^{+0.1}_{-0.1}$ & $0.10^{+0.01}_{-0.01}$ &     C \\
J083856.23+125631.0 & 1739992 &   $169.5^{+0.1}_{-0.1}$ & $17.0^{+0.1}_{-0.1}$ & $0.13^{+0.01}_{-0.01}$ &   $16.1^{+0.1}_{-0.1}$ &  $1.74^{+0.05}_{-0.05}$ &    $58956.9^{+1.1}_{-1.2}$ & $0.08^{+0.01}_{-0.01}$ &     C \\
J084407.61+121642.8 & 1765467 &    $36.2^{+0.1}_{-0.1}$ & $19.0^{+0.2}_{-0.5}$ & $0.04^{+0.01}_{-0.01}$ &   $-5.8^{+0.5}_{-0.2}$ &  $0.81^{+0.06}_{-0.07}$ &    $58136.9^{+0.2}_{-0.4}$ & $0.02^{+0.01}_{-0.01}$ &     C \\
J084607.74+120652.5 & 1765528 &   $265.4^{+0.1}_{-0.1}$ & $15.9^{+0.1}_{-0.1}$ & $0.29^{+0.01}_{-0.01}$ &   $85.1^{+0.1}_{-0.1}$ &  $0.43^{+0.02}_{-0.02}$ &    $59298.3^{+0.5}_{-0.5}$ & $0.09^{+0.01}_{-0.01}$ &     C \\
J084643.40+101907.1 & 1765998 &   $299.0^{+0.8}_{-0.8}$ & $19.0^{+0.3}_{-0.2}$ & $0.12^{+0.02}_{-0.02}$ &  $-59.3^{+0.3}_{-0.3}$ &  $0.30^{+0.08}_{-0.09}$ &    $58937.9^{+3.8}_{-3.8}$ & $0.20^{+0.01}_{-0.01}$ &     C \\
J084437.03+194239.1 & 1766734 &   $143.4^{+0.1}_{-0.1}$ & $21.1^{+0.2}_{-0.2}$ & $0.21^{+0.01}_{-0.02}$ &   $33.0^{+0.2}_{-0.2}$ &  $0.51^{+0.03}_{-0.03}$ &    $59529.7^{+0.6}_{-0.7}$ & $0.13^{+0.01}_{-0.01}$ &     C \\
J084743.29+135618.3 & 1767432 &   $170.8^{+0.1}_{-0.1}$ & $13.1^{+0.1}_{-0.2}$ & $0.02^{+0.01}_{-0.01}$ &   $39.3^{+0.1}_{-0.1}$ &  $1.15^{+0.01}_{-0.02}$ &    $59357.1^{+0.2}_{-0.4}$ & $0.04^{+0.01}_{-0.01}$ &     C \\
J084925.85+111657.0 & 1780702 &   $114.0^{+0.1}_{-0.1}$ & $19.5^{+0.1}_{-0.1}$ & $0.08^{+0.01}_{-0.01}$ &    $2.3^{+0.1}_{-0.1}$ &  $1.66^{+0.01}_{-0.01}$ &    $59264.6^{+0.1}_{-0.1}$ & $0.09^{+0.01}_{-0.01}$ &     C \\
J085026.98+114831.3 & 1780850 &    $62.8^{+0.1}_{-0.1}$ & $26.6^{+0.2}_{-0.2}$ & $0.38^{+0.01}_{-0.01}$ &   $29.3^{+0.1}_{-0.1}$ & $-1.77^{+0.01}_{-0.01}$ &    $59562.8^{+0.1}_{-0.1}$ & $0.10^{+0.01}_{-0.01}$ &     C \\
J085059.23+115636.8 & 1780945 &   $200.2^{+0.2}_{-0.2}$ & $11.8^{+0.1}_{-0.1}$ & $0.10^{+0.01}_{-0.01}$ &   $36.9^{+0.1}_{-0.1}$ &  $0.59^{+0.02}_{-0.02}$ &    $59633.9^{+0.1}_{-0.1}$ & $0.03^{+0.01}_{-0.01}$ &     C \\
J084901.68+125128.3 & 1782464 &   $845.6^{+1.9}_{-1.9}$ &  $7.6^{+0.1}_{-0.1}$ & $0.23^{+0.02}_{-0.02}$ &  $105.8^{+0.1}_{-0.1}$ & $-1.93^{+0.04}_{-0.05}$ &    $59039.5^{+4.0}_{-6.0}$ & $0.03^{+0.01}_{-0.01}$ &     C \\
J084940.47+132958.6 & 1782568 &    $67.0^{+0.1}_{-0.1}$ & $24.9^{+0.1}_{-0.1}$ & $0.23^{+0.01}_{-0.01}$ &  $-72.7^{+0.1}_{-0.1}$ &  $0.83^{+0.02}_{-0.02}$ &    $59275.3^{+0.2}_{-0.2}$ & $0.10^{+0.01}_{-0.01}$ &     C \\
J085424.03+103016.8 & 1796140 &    $33.4^{+0.1}_{-0.1}$ & $26.8^{+0.2}_{-3.3}$ & $0.38^{+0.05}_{-0.01}$ &   $30.1^{+0.2}_{-1.7}$ &  $1.34^{+0.03}_{-0.68}$ &    $59252.2^{+0.1}_{-2.3}$ & $0.05^{+0.01}_{-0.02}$ &     C \\
J085418.55+110038.2 & 1796272 &    $64.9^{+0.1}_{-0.1}$ & $22.6^{+0.2}_{-0.2}$ & $0.37^{+0.01}_{-0.01}$ &   $11.5^{+0.1}_{-0.1}$ &  $2.41^{+0.02}_{-0.02}$ &    $58145.8^{+0.2}_{-0.2}$ & $0.06^{+0.01}_{-0.01}$ &     C \\
J085413.00+125553.1 & 1797901 &   $140.9^{+0.1}_{-0.1}$ & $13.7^{+0.2}_{-0.2}$ & $0.14^{+0.01}_{-0.01}$ &   $13.9^{+0.1}_{-0.1}$ & $-0.97^{+0.02}_{-0.02}$ &    $58046.0^{+0.1}_{-0.1}$ & $0.04^{+0.01}_{-0.01}$ &     C \\
J090321.19+422908.2 & 1827207 &    $89.8^{+0.1}_{-0.1}$ & $20.6^{+0.1}_{-0.1}$ & $0.21^{+0.01}_{-0.01}$ &    $5.5^{+0.1}_{-0.1}$ &  $0.89^{+0.02}_{-0.02}$ &    $59683.4^{+0.2}_{-0.3}$ & $0.07^{+0.01}_{-0.01}$ &     C \\
J090658.70+435707.3 & 1843282 &    $52.2^{+0.1}_{-0.1}$ & $26.2^{+0.1}_{-0.2}$ & $0.40^{+0.01}_{-0.01}$ &   $25.3^{+0.1}_{-0.1}$ & $-2.25^{+0.01}_{-0.01}$ &    $59592.2^{+0.1}_{-0.1}$ & $0.07^{+0.01}_{-0.01}$ &     C \\
J091705.56+415450.3 & 1890590 &   $110.5^{+0.1}_{-0.1}$ & $24.7^{+0.1}_{-0.1}$ & $0.59^{+0.01}_{-0.01}$ &    $0.9^{+0.1}_{-0.1}$ & $-0.72^{+0.01}_{-0.01}$ &    $59242.3^{+0.1}_{-0.1}$ & $0.09^{+0.01}_{-0.01}$ &     C \\
J104129.64+090143.6 & 2132701 &   $187.2^{+0.1}_{-0.1}$ & $18.8^{+0.1}_{-0.1}$ & $0.40^{+0.01}_{-0.01}$ &    $6.7^{+0.1}_{-0.1}$ & $-1.18^{+0.02}_{-0.02}$ &    $59277.0^{+0.3}_{-0.3}$ & $0.10^{+0.01}_{-0.01}$ &     C \\
J104150.69+095125.1 & 2132775 &    $77.1^{+0.1}_{-0.1}$ & $25.7^{+0.1}_{-0.1}$ & $0.31^{+0.01}_{-0.01}$ &   $83.7^{+0.1}_{-0.1}$ &  $0.53^{+0.02}_{-0.02}$ &    $60332.6^{+0.2}_{-0.1}$ & $0.11^{+0.01}_{-0.01}$ &     C \\
J104507.39+102053.5 & 2143187 &   $202.1^{+0.1}_{-0.1}$ & $14.5^{+0.1}_{-0.1}$ & $0.17^{+0.01}_{-0.01}$ &   $29.4^{+0.1}_{-0.1}$ & $-3.03^{+0.02}_{-0.02}$ &    $59899.0^{+0.4}_{-0.5}$ & $0.06^{+0.01}_{-0.01}$ &     C \\
J104542.58+084305.5 & 2143641 &   $146.2^{+0.1}_{-0.1}$ & $18.8^{+0.1}_{-0.1}$ & $0.03^{+0.01}_{-0.01}$ &   $41.0^{+0.1}_{-0.1}$ & $-2.26^{+0.11}_{-0.07}$ &    $59616.8^{+2.6}_{-1.5}$ & $0.10^{+0.01}_{-0.01}$ &     C \\
J104529.11+410936.7 & 2144691 &   $140.1^{+0.1}_{-0.1}$ & $13.4^{+0.1}_{-0.1}$ & $0.34^{+0.01}_{-0.01}$ &  $-37.8^{+0.1}_{-0.1}$ &  $1.06^{+0.02}_{-0.03}$ &    $59189.3^{+0.4}_{-0.4}$ & $0.03^{+0.01}_{-0.01}$ &     C \\
J104912.62+405323.7 & 2154869 &    $98.1^{+0.1}_{-0.1}$ & $23.7^{+0.2}_{-0.2}$ & $0.45^{+0.01}_{-0.01}$ &  $-38.5^{+0.1}_{-0.1}$ &  $2.28^{+0.02}_{-0.02}$ &    $58506.1^{+0.2}_{-0.2}$ & $0.09^{+0.01}_{-0.01}$ &     C \\
J112249.40+024806.9 & 2231766 &   $185.8^{+0.1}_{-0.1}$ & $15.5^{+0.1}_{-0.1}$ & $0.12^{+0.01}_{-0.01}$ &   $29.7^{+0.1}_{-0.1}$ & $-3.14^{+0.01}_{-0.01}$ &    $59209.2^{+0.3}_{-0.2}$ & $0.07^{+0.01}_{-0.01}$ &     C \\
J112001.97+033213.6 & 2232132 &   $213.7^{+0.2}_{-0.2}$ & $17.6^{+0.2}_{-0.2}$ & $0.44^{+0.01}_{-0.01}$ &   $48.4^{+0.1}_{-0.1}$ & $-2.40^{+0.02}_{-0.02}$ &    $59219.2^{+0.7}_{-0.6}$ & $0.08^{+0.01}_{-0.01}$ &     C \\
J114552.77+351726.6 & 2301369 &   $138.9^{+0.1}_{-0.1}$ & $19.1^{+0.1}_{-0.1}$ & $0.22^{+0.01}_{-0.01}$ &   $35.0^{+0.1}_{-0.1}$ &  $0.12^{+0.02}_{-0.02}$ &    $60087.3^{+0.3}_{-0.4}$ & $0.09^{+0.01}_{-0.01}$ &     C \\
J114849.82+344617.0 & 2311068 &    $61.8^{+0.1}_{-0.1}$ & $25.8^{+0.1}_{-0.1}$ & $0.49^{+0.01}_{-0.01}$ &  $-10.3^{+0.1}_{-0.1}$ &  $0.48^{+0.01}_{-0.01}$ &    $58549.3^{+0.1}_{-0.1}$ & $0.07^{+0.01}_{-0.01}$ &     C \\
J115407.72+335630.1 & 2320508 &   $189.9^{+0.1}_{-0.1}$ & $15.9^{+0.1}_{-0.1}$ & $0.35^{+0.01}_{-0.01}$ &   $50.8^{+0.1}_{-0.1}$ &  $0.36^{+0.02}_{-0.02}$ &    $58474.8^{+0.5}_{-0.4}$ & $0.06^{+0.01}_{-0.01}$ &     C \\
J115806.74+352745.0 & 2327442 &   $246.1^{+0.4}_{-0.4}$ & $14.1^{+0.1}_{-0.1}$ & $0.14^{+0.01}_{-0.01}$ &  $-41.5^{+0.1}_{-0.1}$ &  $2.01^{+0.04}_{-0.04}$ &    $58988.6^{+1.4}_{-1.3}$ & $0.07^{+0.01}_{-0.01}$ &     C \\
J115641.99+362234.5 & 2327619 &    $97.1^{+0.1}_{-0.1}$ & $22.1^{+0.2}_{-0.2}$ & $0.45^{+0.01}_{-0.01}$ &  $-36.7^{+0.2}_{-0.1}$ &  $0.48^{+0.02}_{-0.02}$ &    $58950.0^{+0.2}_{-0.2}$ & $0.08^{+0.01}_{-0.01}$ &     C \\
J120015.62+342700.2 & 2334691 &   $319.3^{+0.6}_{-0.7}$ & $12.0^{+0.2}_{-0.2}$ & $0.05^{+0.02}_{-0.02}$ &    $4.5^{+0.2}_{-0.2}$ & $-0.45^{+0.20}_{-0.16}$ &   $58982.5^{+10.1}_{-8.0}$ & $0.06^{+0.01}_{-0.01}$ &     C \\
J132900.05+513023.8 & 2537667 &    $45.6^{+0.1}_{-0.1}$ & $31.2^{+0.3}_{-2.1}$ & $0.07^{+0.01}_{-0.01}$ &  $-57.8^{+0.3}_{-0.2}$ &  $0.04^{+0.02}_{-0.01}$ &    $59287.3^{+0.2}_{-0.2}$ & $0.14^{+0.01}_{-0.03}$ &     C \\
J140746.12+454135.5 & 2624887 &   $502.2^{+0.7}_{-0.6}$ & $13.3^{+0.1}_{-0.1}$ & $0.40^{+0.01}_{-0.01}$ & $-190.0^{+0.1}_{-0.1}$ & $-2.91^{+0.01}_{-0.01}$ &    $58946.0^{+0.6}_{-0.5}$ & $0.09^{+0.01}_{-0.01}$ &     C \\
J142728.19+453124.5 & 2662589 &   $519.5^{+1.0}_{-0.9}$ & $11.4^{+0.1}_{-0.1}$ & $0.18^{+0.01}_{-0.01}$ &  $-31.2^{+0.1}_{-0.1}$ &  $1.69^{+0.03}_{-0.04}$ &    $59524.9^{+3.0}_{-3.1}$ & $0.07^{+0.01}_{-0.01}$ &     C \\
J230706.97+352452.3 & 3131875 &    $11.2^{+0.1}_{-0.1}$ & $42.6^{+0.2}_{-0.1}$ & $0.08^{+0.01}_{-0.01}$ &   $-6.7^{+0.1}_{-0.3}$ & $-2.58^{+0.03}_{-0.03}$ &    $59869.8^{+0.1}_{-0.1}$ & $0.09^{+0.01}_{-0.01}$ &     C \\

\label{tab:orbit}
\end{longtable}

\begin{minipage}{\textwidth}
    \vspace{-2pt} 
    \footnotesize
    \textbf{Notes.} \\
    Column (1): Source designation in LAMOST catalog; 
    Column (2): Source ID in our sample; 
    Column (3): Orbital period; 
    Column (4): Semi-amplitude of RV curve; 
    Column (5): Eccentricity; 
    Column (6): Systemic velocity; 
    Column (7): Argument of periastron; 
    Column (8): Time of periastron passage in \texttt{lmjd}; 
    Column (9): Mass function calculated based on Columns (3), (4), and (5); 
    Column (10): Candidates classification. \\
$\ast$ The deficiency of candidates in the RA range of $14^{\rm h}$ to $23^{\rm h}$ is attributed to the seasonal observation constraints of LAMOST. In this celestial region, there is a sharp decrease in the number of targets with sufficient observation epochs to satisfy our minimum requirements..
\end{minipage}

\clearpage
\begin{longtable}{@{\extracolsep{\fill}}rrrrrrrr}
\caption{Stellar Parameters for the Candidates}\\
\toprule
\hline
     SID & $T_{\rm eff}$ & $\log g$ & [Fe/H] & $R_1$ & $M_1$  & $M_{2}^\mathrm{min}$ & Class \\
      & (K) &  &  & ($R_{\odot}$) & ($M_{\odot}$) & ($M_{\odot}$) & \\
     (1) & (2) & (3) & (4) & (5) & (6) & (7) & (8) \\
\midrule
\endfirsthead
\caption[]{Stellar Parameters — Continued} \\
\toprule
     SID & $T_{\rm eff}$ & $\log g$ & [Fe/H] & $R_1$ & $M_1$  & $M_{2}^\mathrm{min}$ & Class \\
      & (K) &  &  & ($R_{\odot}$) & ($M_{\odot}$) & ($M_{\odot}$) & \\
     (1) & (2) & (3) & (4) & (5) & (6) & (7) & (8) \\
\midrule
\endhead
\midrule
\multicolumn{8}{r}{{Continued on next page}} \\
\midrule
\endfoot

\bottomrule
\endlastfoot
1098054 &   $6181^{+29}_{-35}$ & $4.22^{+0.04}_{-0.04}$ & $-0.27^{+0.05}_{-0.05}$ &    $2.03^{+0.08}_{-0.06}$ & $1.19^{+0.04}_{-0.05}$ & $1.33^{+0.05}_{-0.06}$ &     A \\
1753900 & $5872^{+174}_{-122}$ & $4.01^{+1.57}_{-0.48}$ & $-0.42^{+0.31}_{-0.31}$ &    $1.55^{+0.27}_{-0.15}$ & $0.90^{+0.13}_{-0.08}$ & $0.79^{+0.06}_{-0.04}$ &     A \\
1767926 &   $4593^{+11}_{-13}$ & $5.00^{+0.02}_{-0.03}$ & $-0.57^{+0.02}_{-0.01}$ &    $1.87^{+0.35}_{-0.09}$ & $0.81^{+0.06}_{-0.06}$ & $0.82^{+0.06}_{-0.08}$ &     A \\
1795960 &   $4936^{+14}_{-16}$ & $4.70^{+0.07}_{-0.04}$ & $-0.06^{+0.02}_{-0.02}$ &    $0.87^{+0.06}_{-0.05}$ & $0.89^{+0.08}_{-0.07}$ & $1.10^{+0.05}_{-0.05}$ &     A \\
 441454 &   $4380^{+12}_{-12}$ & $2.01^{+0.02}_{-0.02}$ & $-1.08^{+0.02}_{-0.02}$ &   $22.04^{+1.12}_{-0.99}$ & $0.81^{+0.04}_{-0.03}$ & $1.24^{+0.67}_{-2.75}$ &     B \\
 454174 & $5742^{+250}_{-339}$ & $4.33^{+0.61}_{-0.63}$ & $-0.23^{+0.18}_{-0.18}$ &    $5.97^{+0.40}_{-0.30}$ & $1.41^{+0.48}_{-0.43}$ & $1.07^{+0.18}_{-0.18}$ &     B \\
 454319 &   $4602^{+12}_{-11}$ & $3.26^{+0.04}_{-0.03}$ & $-0.63^{+0.01}_{-0.01}$ &    $9.21^{+0.26}_{-0.25}$ & $0.84^{+0.06}_{-0.02}$ & $0.80^{+0.03}_{-0.01}$ &     B \\
 488064 & $4140^{+294}_{-133}$ & $3.78^{+0.37}_{-0.18}$ & $-0.15^{+0.18}_{-0.14}$ &   $53.08^{+4.62}_{-3.87}$ & $1.88^{+0.10}_{-0.23}$ & $2.72^{+0.08}_{-3.39}$ &     B \\
 782766 & $4803^{+426}_{-326}$ & $4.37^{+0.90}_{-0.67}$ & $-0.18^{+0.30}_{-0.24}$ &   $13.71^{+2.15}_{-1.39}$ & $1.64^{+0.25}_{-0.62}$ & $1.24^{+0.11}_{-0.27}$ &     B \\
 865354 &   $5149^{+32}_{-34}$ & $3.61^{+0.05}_{-0.07}$ & $-0.71^{+0.04}_{-0.03}$ &    $9.84^{+1.06}_{-0.54}$ & $0.90^{+0.47}_{-0.12}$ & $1.02^{+0.24}_{-0.07}$ &     B \\
 877473 & $4573^{+242}_{-291}$ & $4.07^{+0.67}_{-0.47}$ & $-0.26^{+0.23}_{-0.23}$ &   $17.82^{+1.93}_{-1.67}$ & $2.39^{+0.27}_{-1.30}$ & $2.14^{+0.19}_{-0.62}$ &     B \\
 878211 & $4950^{+690}_{-319}$ & $4.65^{+0.62}_{-0.79}$ & $-0.16^{+0.22}_{-0.27}$ &   $15.09^{+3.03}_{-1.40}$ & $1.08^{+0.19}_{-0.17}$ & $1.33^{+0.11}_{-0.10}$ &     B \\
1890750 &   $4797^{+16}_{-19}$ & $3.71^{+0.03}_{-0.04}$ & $-1.00^{+0.01}_{-0.01}$ &   $11.20^{+1.36}_{-0.51}$ & $0.77^{+0.04}_{-0.02}$ & $0.78^{+0.02}_{-0.01}$ &     B \\
 109269 &   $6069^{+74}_{-73}$ & $4.34^{+0.77}_{-0.59}$ & $-0.22^{+0.15}_{-0.11}$ &    $4.69^{+1.42}_{-1.13}$ & $2.05^{+0.37}_{-0.40}$ & $1.20^{+0.12}_{-0.14}$ &     C \\
 422741 & $3933^{+147}_{-105}$ & $3.65^{+0.46}_{-0.14}$ & $-0.14^{+0.33}_{-0.21}$ &   $40.68^{+4.98}_{-2.78}$ & $1.17^{+0.19}_{-0.25}$ & $0.77^{+0.07}_{-0.10}$ &     C \\
 453063 & $4196^{+247}_{-213}$ & $4.18^{+0.46}_{-0.39}$ & $-0.14^{+0.15}_{-0.15}$ &   $17.21^{+2.22}_{-1.27}$ & $1.04^{+0.14}_{-0.04}$ & $0.68^{+0.05}_{-0.03}$ &     C \\
 576683 & $4299^{+200}_{-170}$ & $3.97^{+0.48}_{-0.27}$ & $-0.24^{+0.18}_{-0.17}$ &   $39.07^{+2.66}_{-3.09}$ & $3.79^{+0.33}_{-2.45}$ & $2.65^{+0.15}_{-1.06}$ &     C \\
 623589 &   $5600^{+15}_{-18}$ & $4.15^{+0.03}_{-0.03}$ &  $0.25^{+0.02}_{-0.02}$ &    $1.27^{+0.06}_{-0.04}$ & $1.03^{+0.03}_{-0.03}$ & $0.59^{+0.01}_{-0.01}$ &     C \\
 686208 &   $5860^{+26}_{-28}$ & $3.38^{+0.03}_{-0.03}$ &  $0.28^{+0.03}_{-0.03}$ &   $10.11^{+0.93}_{-0.62}$ & $2.91^{+0.18}_{-0.75}$ & $1.64^{+0.08}_{-0.24}$ &     C \\
 808439 & $4698^{+397}_{-613}$ & $4.44^{+0.87}_{-0.72}$ & $-0.22^{+0.29}_{-0.22}$ &   $17.65^{+4.37}_{-4.51}$ & $1.94^{+0.49}_{-1.13}$ & $1.18^{+0.17}_{-0.43}$ &     C \\
 824490 & $4822^{+528}_{-233}$ & $4.55^{+0.55}_{-0.51}$ & $-0.12^{+0.19}_{-0.20}$ &   $12.06^{+0.77}_{-0.65}$ & $1.26^{+0.22}_{-0.38}$ & $0.85^{+0.09}_{-0.15}$ &     C \\
 888996 & $4462^{+294}_{-164}$ & $3.96^{+0.60}_{-0.32}$ & $-0.23^{+0.20}_{-0.24}$ &   $24.59^{+3.78}_{-1.84}$ & $2.80^{+0.37}_{-1.62}$ & $1.96^{+0.15}_{-0.69}$ &     C \\
 889042 &   $5954^{+59}_{-79}$ & $3.24^{+0.02}_{-0.03}$ &  $0.66^{+0.05}_{-0.02}$ &  $37.89^{+0.84}_{-11.43}$ & $6.08^{+0.59}_{-1.75}$ & $3.59^{+0.21}_{-0.60}$ &     C \\
1047481 &     $5005^{+9}_{-9}$ & $3.84^{+0.03}_{-0.03}$ &  $0.06^{+0.01}_{-0.01}$ &    $3.17^{+0.11}_{-0.09}$ & $1.36^{+0.08}_{-0.09}$ & $0.71^{+0.03}_{-0.03}$ &     C \\
1063997 &   $5607^{+22}_{-19}$ & $3.24^{+0.03}_{-0.03}$ & $-0.25^{+0.02}_{-0.02}$ &   $14.42^{+1.03}_{-0.99}$ & $4.89^{+0.44}_{-0.44}$ & $2.44^{+0.13}_{-0.13}$ &     C \\
1080054 &   $5825^{+33}_{-34}$ & $3.63^{+0.03}_{-0.03}$ &  $0.28^{+0.04}_{-0.04}$ &   $10.32^{+0.65}_{-0.63}$ & $2.72^{+0.18}_{-0.18}$ & $1.63^{+0.06}_{-0.06}$ &     C \\
1167929 &   $4742^{+15}_{-13}$ & $3.22^{+0.05}_{-0.05}$ & $-0.10^{+0.02}_{-0.02}$ &   $16.37^{+1.46}_{-1.05}$ & $2.52^{+0.24}_{-0.32}$ & $1.78^{+0.11}_{-0.13}$ &     C \\
1171404 &   $5639^{+34}_{-39}$ & $3.63^{+0.04}_{-0.04}$ & $-0.37^{+0.04}_{-0.04}$ &    $8.30^{+0.51}_{-0.44}$ & $2.35^{+0.14}_{-0.19}$ & $1.64^{+0.06}_{-0.08}$ &     C \\
1186400 &   $5930^{+33}_{-34}$ & $2.65^{+0.06}_{-0.06}$ &  $0.28^{+0.06}_{-0.05}$ &   $22.49^{+3.38}_{-2.72}$ & $3.93^{+0.93}_{-0.70}$ & $1.52^{+0.20}_{-0.17}$ &     C \\
1206050 & $4294^{+199}_{-274}$ & $4.29^{+0.66}_{-0.53}$ & $-0.10^{+0.27}_{-0.26}$ &   $42.42^{+8.54}_{-5.45}$ & $2.65^{+0.21}_{-1.22}$ & $1.54^{+0.08}_{-0.44}$ &     C \\
1243191 & $4702^{+129}_{-413}$ & $4.53^{+0.65}_{-0.92}$ & $-0.16^{+0.28}_{-0.27}$ &   $13.64^{+1.67}_{-1.16}$ & $2.03^{+0.22}_{-0.59}$ & $1.34^{+0.09}_{-0.22}$ &     C \\
1299572 &  $4311^{+305}_{-80}$ & $3.90^{+0.65}_{-0.38}$ & $-0.08^{+0.20}_{-0.41}$ &   $12.07^{+4.38}_{-0.95}$ & $2.44^{+0.37}_{-0.75}$ & $0.91^{+0.10}_{-0.18}$ &     C \\
1316116 &  $4461^{+56}_{-135}$ & $4.88^{+0.49}_{-1.08}$ &  $0.01^{+0.26}_{-0.40}$ &    $0.79^{+0.04}_{-0.03}$ & $0.82^{+0.04}_{-0.06}$ & $0.45^{+0.02}_{-0.02}$ &     C \\
1592874 &   $6029^{+46}_{-48}$ & $4.19^{+0.04}_{-0.04}$ & $-0.45^{+0.06}_{-0.05}$ &    $1.75^{+0.11}_{-0.05}$ & $0.99^{+0.07}_{-0.08}$ & $0.45^{+0.02}_{-0.02}$ &     C \\
1606805 &   $6011^{+50}_{-80}$ & $4.65^{+0.10}_{-0.07}$ & $-1.03^{+0.08}_{-0.09}$ &    $1.52^{+0.14}_{-0.13}$ & $0.81^{+0.07}_{-0.02}$ & $0.47^{+0.02}_{-0.01}$ &     C \\
1607121 &     $4750^{+7}_{-5}$ & $3.40^{+0.06}_{-0.04}$ & $-0.25^{+0.02}_{-0.01}$ &    $5.61^{+0.60}_{-0.16}$ & $1.33^{+0.06}_{-0.12}$ & $1.00^{+0.03}_{-0.05}$ &     C \\
1635668 &   $5418^{+31}_{-38}$ & $4.34^{+0.05}_{-0.05}$ & $-0.53^{+0.03}_{-0.05}$ &    $2.55^{+0.26}_{-0.07}$ & $1.05^{+0.12}_{-0.11}$ & $0.63^{+0.04}_{-0.04}$ &     C \\
1651164 &   $5726^{+40}_{-67}$ & $4.38^{+0.14}_{-0.06}$ & $-0.00^{+0.05}_{-0.09}$ &    $1.46^{+0.17}_{-0.05}$ & $0.99^{+0.06}_{-0.07}$ & $0.49^{+0.02}_{-0.02}$ &     C \\
1738351 &   $5758^{+74}_{-42}$ & $4.14^{+0.05}_{-0.03}$ &  $0.38^{+0.05}_{-0.05}$ &    $1.19^{+0.05}_{-0.06}$ & $1.05^{+0.06}_{-0.06}$ & $0.65^{+0.02}_{-0.02}$ &     C \\
1739992 &   $6018^{+39}_{-40}$ & $4.34^{+0.04}_{-0.04}$ & $-0.46^{+0.04}_{-0.05}$ &    $1.64^{+0.08}_{-0.04}$ & $0.97^{+0.04}_{-0.04}$ & $0.58^{+0.02}_{-0.02}$ &     C \\
1765467 &   $4817^{+11}_{-12}$ & $4.27^{+0.03}_{-0.04}$ & $-0.81^{+0.02}_{-0.02}$ &    $6.21^{+0.22}_{-0.21}$ & $0.82^{+0.02}_{-0.02}$ & $0.32^{+0.01}_{-0.01}$ &     C \\
1765528 &   $5483^{+19}_{-21}$ & $3.94^{+0.03}_{-0.03}$ &  $0.25^{+0.02}_{-0.02}$ &    $2.09^{+0.04}_{-0.12}$ & $1.15^{+0.13}_{-0.10}$ & $0.68^{+0.04}_{-0.04}$ &     C \\
1765998 &   $4302^{+79}_{-56}$ & $5.09^{+0.55}_{-0.59}$ & $-0.10^{+0.32}_{-0.34}$ & $78.78^{+10.26}_{-23.21}$ & $4.58^{+0.28}_{-1.58}$ & $2.06^{+0.10}_{-0.43}$ &     C \\
1766734 &   $6145^{+26}_{-32}$ & $4.37^{+0.04}_{-0.04}$ &  $0.07^{+0.03}_{-0.03}$ &    $1.20^{+0.11}_{-0.05}$ & $1.15^{+0.03}_{-0.04}$ & $0.78^{+0.02}_{-0.02}$ &     C \\
1767432 &   $5871^{+44}_{-50}$ & $4.84^{+0.06}_{-0.06}$ & $-0.47^{+0.05}_{-0.05}$ &    $1.49^{+0.07}_{-0.05}$ & $0.89^{+0.04}_{-0.04}$ & $0.40^{+0.02}_{-0.02}$ &     C \\
1780702 &   $5893^{+35}_{-32}$ & $4.29^{+0.03}_{-0.03}$ &  $0.35^{+0.04}_{-0.05}$ &    $1.53^{+0.04}_{-0.04}$ & $1.25^{+0.04}_{-0.04}$ & $0.68^{+0.02}_{-0.02}$ &     C \\
1780850 &   $5999^{+37}_{-33}$ & $4.13^{+0.04}_{-0.04}$ &  $0.00^{+0.04}_{-0.05}$ &    $1.64^{+0.05}_{-0.04}$ & $1.13^{+0.11}_{-0.05}$ & $0.68^{+0.04}_{-0.02}$ &     C \\
1780945 &   $6207^{+28}_{-27}$ & $4.19^{+0.03}_{-0.03}$ &  $0.16^{+0.04}_{-0.04}$ &    $1.57^{+0.07}_{-0.06}$ & $1.32^{+0.06}_{-0.07}$ & $0.47^{+0.02}_{-0.02}$ &     C \\
1782464 &   $4927^{+46}_{-39}$ & $4.05^{+0.45}_{-0.41}$ & $-0.31^{+0.18}_{-0.21}$ &   $12.91^{+1.82}_{-1.90}$ & $1.37^{+0.38}_{-0.25}$ & $0.49^{+0.08}_{-0.06}$ &     C \\
1782568 &   $5460^{+31}_{-90}$ & $4.99^{+0.01}_{-0.03}$ & $-1.06^{+0.03}_{-0.03}$ &    $1.29^{+0.09}_{-0.02}$ & $0.98^{+0.03}_{-0.02}$ & $0.63^{+0.01}_{-0.01}$ &     C \\
1796140 &   $6042^{+41}_{-40}$ & $4.22^{+0.04}_{-0.04}$ &  $0.17^{+0.06}_{-0.07}$ &    $1.96^{+0.08}_{-0.07}$ & $1.39^{+0.06}_{-0.19}$ & $0.55^{+0.05}_{-0.08}$ &     C \\
1796272 &   $6023^{+37}_{-39}$ & $4.38^{+0.04}_{-0.05}$ & $-0.45^{+0.05}_{-0.05}$ &    $1.14^{+0.04}_{-0.02}$ & $0.89^{+0.04}_{-0.04}$ & $0.48^{+0.02}_{-0.02}$ &     C \\
1797901 &   $5659^{+85}_{-73}$ & $4.18^{+0.65}_{-0.56}$ & $-0.09^{+0.21}_{-0.26}$ &    $1.86^{+0.19}_{-0.10}$ & $1.05^{+0.11}_{-0.11}$ & $0.43^{+0.03}_{-0.03}$ &     C \\
1827207 &   $5716^{+38}_{-36}$ & $4.25^{+0.05}_{-0.04}$ &  $0.02^{+0.04}_{-0.05}$ &    $1.56^{+0.09}_{-0.07}$ & $1.01^{+0.05}_{-0.04}$ & $0.57^{+0.02}_{-0.02}$ &     C \\
1843282 &   $5949^{+33}_{-31}$ & $4.63^{+0.03}_{-0.03}$ & $-0.39^{+0.04}_{-0.04}$ &    $1.19^{+0.04}_{-0.03}$ & $0.87^{+0.04}_{-0.03}$ & $0.52^{+0.02}_{-0.01}$ &     C \\
1890590 &   $5639^{+30}_{-29}$ & $4.43^{+0.05}_{-0.04}$ & $-0.24^{+0.03}_{-0.03}$ &    $1.57^{+0.04}_{-0.03}$ & $0.91^{+0.03}_{-0.02}$ & $0.58^{+0.01}_{-0.01}$ &     C \\
2132701 &   $4699^{+44}_{-43}$ & $4.53^{+0.72}_{-0.75}$ & $-0.19^{+0.21}_{-0.21}$ &    $5.64^{+0.93}_{-0.28}$ & $2.19^{+0.08}_{-0.33}$ & $0.99^{+0.03}_{-0.09}$ &     C \\
2132775 &   $5035^{+62}_{-51}$ & $4.15^{+0.79}_{-0.59}$ & $-0.44^{+0.35}_{-0.26}$ &    $6.32^{+0.84}_{-0.82}$ & $0.83^{+0.13}_{-0.05}$ & $0.62^{+0.05}_{-0.02}$ &     C \\
2143187 &   $5034^{+17}_{-20}$ & $4.49^{+0.03}_{-0.03}$ & $-0.01^{+0.02}_{-0.02}$ &    $0.88^{+0.03}_{-0.02}$ & $0.87^{+0.02}_{-0.07}$ & $0.48^{+0.01}_{-0.02}$ &     C \\
2143641 &   $4838^{+22}_{-25}$ & $4.89^{+0.10}_{-0.08}$ & $-0.44^{+0.02}_{-0.03}$ &    $0.88^{+0.06}_{-0.03}$ & $0.77^{+0.05}_{-0.04}$ & $0.56^{+0.02}_{-0.02}$ &     C \\
2144691 &   $5386^{+32}_{-31}$ & $4.36^{+0.06}_{-0.05}$ & $-0.29^{+0.05}_{-0.04}$ &    $1.99^{+0.15}_{-0.11}$ & $0.94^{+0.10}_{-0.07}$ & $0.36^{+0.03}_{-0.02}$ &     C \\
2154869 &   $5638^{+50}_{-54}$ & $4.52^{+0.05}_{-0.04}$ & $-0.41^{+0.14}_{-0.14}$ &    $1.58^{+0.06}_{-0.07}$ & $1.23^{+0.07}_{-0.08}$ & $0.71^{+0.03}_{-0.03}$ &     C \\
2231766 &   $6003^{+42}_{-47}$ & $4.23^{+0.03}_{-0.03}$ &  $0.18^{+0.05}_{-0.06}$ &    $1.82^{+0.07}_{-0.06}$ & $1.35^{+0.06}_{-0.17}$ & $0.65^{+0.02}_{-0.05}$ &     C \\
2232132 &   $5301^{+41}_{-45}$ & $4.90^{+0.05}_{-0.06}$ & $-0.62^{+0.03}_{-0.03}$ &    $1.19^{+0.07}_{-0.04}$ & $1.00^{+0.07}_{-0.06}$ & $0.60^{+0.03}_{-0.02}$ &     C \\
2301369 &   $5943^{+26}_{-34}$ & $4.26^{+0.03}_{-0.04}$ & $-0.49^{+0.03}_{-0.03}$ &    $2.11^{+0.07}_{-0.05}$ & $1.04^{+0.04}_{-0.03}$ & $0.64^{+0.02}_{-0.02}$ &     C \\
2311068 &   $5253^{+18}_{-23}$ & $4.42^{+0.05}_{-0.05}$ & $-0.06^{+0.02}_{-0.03}$ &    $1.14^{+0.05}_{-0.03}$ & $1.04^{+0.06}_{-0.05}$ & $0.57^{+0.02}_{-0.02}$ &     C \\
2320508 &   $5214^{+41}_{-50}$ & $4.95^{+0.05}_{-0.04}$ & $-1.58^{+0.03}_{-0.02}$ &    $1.85^{+0.12}_{-0.10}$ & $0.97^{+0.06}_{-0.06}$ & $0.52^{+0.02}_{-0.02}$ &     C \\
2327442 &   $5888^{+38}_{-41}$ & $4.55^{+0.05}_{-0.04}$ & $-0.14^{+0.05}_{-0.05}$ &    $1.07^{+0.12}_{-0.05}$ & $0.95^{+0.05}_{-0.04}$ & $0.53^{+0.02}_{-0.02}$ &     C \\
2327619 &   $6009^{+55}_{-99}$ & $4.52^{+0.09}_{-0.07}$ & $-0.26^{+0.05}_{-0.06}$ &    $1.10^{+0.05}_{-0.05}$ & $0.91^{+0.07}_{-0.06}$ & $0.54^{+0.03}_{-0.02}$ &     C \\
2334691 &   $4701^{+15}_{-10}$ & $4.29^{+0.08}_{-0.08}$ &  $0.02^{+0.02}_{-0.02}$ &    $0.81^{+0.13}_{-0.02}$ & $0.74^{+0.03}_{-0.02}$ & $0.42^{+0.01}_{-0.01}$ &     C \\
2537667 &   $4746^{+11}_{-10}$ & $3.39^{+0.04}_{-0.05}$ & $-0.26^{+0.02}_{-0.02}$ &    $9.24^{+0.47}_{-0.28}$ & $1.25^{+0.10}_{-0.08}$ & $0.83^{+0.05}_{-0.07}$ &     C \\
2624887 &   $4761^{+36}_{-32}$ & $3.92^{+0.81}_{-0.29}$ & $-0.32^{+0.20}_{-0.19}$ &    $9.15^{+0.63}_{-0.37}$ & $1.13^{+0.36}_{-0.27}$ & $0.67^{+0.12}_{-0.10}$ &     C \\
2662589 &   $5000^{+21}_{-17}$ & $4.07^{+0.05}_{-0.06}$ & $-0.37^{+0.02}_{-0.02}$ &    $4.01^{+0.21}_{-0.08}$ & $1.08^{+0.08}_{-0.08}$ & $0.59^{+0.03}_{-0.03}$ &     C \\
3131875 &   $6111^{+37}_{-34}$ & $3.80^{+0.03}_{-0.03}$ &  $0.03^{+0.05}_{-0.05}$ &    $3.01^{+0.10}_{-0.08}$ & $1.53^{+0.08}_{-0.08}$ & $0.77^{+0.03}_{-0.03}$ &     C \\
\label{tab:params}
\end{longtable}

\begin{minipage}{\textwidth}
    \vspace{-2pt} 
    \footnotesize
    \textbf{Notes.} \\
    Column (1): Source ID in our sample; 
    Columns (2) -- (5): Effective temperature, surface gravity, metallicity and stellar radius from SED fitting results; 
    Column (6): Primary mass estimated from isochrone; 
    Column (7): Minimum secondary mass calculated at $i=90^{\circ}$; 
    Column (8): Candidates classification. \\
$\ast$ For some candidates, the uncertainties for $\log g$ and [Fe/H] are reported as less than $0.1$ dex. This is because the results from spectroscopic template matching were adopted as informative priors for the SED fitting. These values only represent the statistical uncertainties under a constrained parameter space. Conversely, for targets where such priors resulted in poor convergence, a non-informative uniform prior was applied, yielding more conservative and substantially larger uncertainties that reflect the full range of physically plausible solutions.
\end{minipage}

\end{document}